\documentclass{article}

\usepackage{PRIMEarxiv}

\usepackage[utf8]{inputenc} 
\usepackage[T1]{fontenc}    
\usepackage{hyperref}       
\usepackage{url}            
\usepackage{booktabs}       
\usepackage{amsfonts}       
\usepackage{nicefrac}       
\usepackage{microtype}      
\usepackage{lipsum}
\usepackage{fancyhdr}       
\usepackage{graphicx}       
\graphicspath{{graphs/}}     
\usepackage{caption}
\usepackage{subcaption}
\usepackage[dvipsnames]{xcolor}
\usepackage{epsfig}
\usepackage{textcomp}
\usepackage{bm}
\usepackage{algorithm}
\usepackage{algorithmic}
\usepackage{dsfont}

\newtheorem{proposition}{Proposition}

\usepackage{arydshln}

 \usepackage{amsmath}
 \usepackage{amssymb}
 \usepackage{epsfig}
\usepackage{rotating}
 \usepackage{url}
 \usepackage{amsfonts}
 \usepackage{mathrsfs}
 \usepackage{textcomp}
 
\usepackage{mathrsfs}
\usepackage{natbib}
\usepackage{nccmath}

\usepackage{enumitem}

\title{A novel approach to assess dynamic treatment regimes embedded in a SMART with an ordinal outcome}

\author{PALASH GHOSH$^{1,2,\ast}$, XIAOXI YAN$^{2}$ and BIBHAS CHAKRABORTY$^{2,3,4}$\\[4pt]
	\textit{$^{1}$Department of Mathematics and School of Health Science \& Technology,
		Indian Institute of Technology Guwahati, India}
	\\[2pt]
		\textit{$^{2}$Centre for Quantitative Medicine, Duke-NUS Medical School, Singapore}
	\\[2pt]
		\textit{$^{3}$Department of Statistics and Data Science, National University of Singapore, Singaporee}
	\\[2pt]
		\textit{$^{4}$Department of Biostatistics and Bioinformatics, Duke University, Durham, NC,USA}
	\\[2pt]
	{*palash.ghosh@iitg.ac.in}}

\begin{document}
\maketitle

\begin{abstract}
	{Sequential multiple assignment randomized trials (SMARTs) are used to construct data-driven optimal intervention strategies for subjects based on their intervention and covariate histories in different branches of health and behavioral sciences where a sequence of interventions is given to a participant. Sequential intervention strategies are often called dynamic treatment regimes (DTR). In the existing literature, the majority of the analysis methodologies for SMART data assume a continuous primary outcome. However, ordinal outcomes are also quite common in clinical practice. In this work, first, we introduce the notion of generalized odds ratio ($GOR$) to compare two DTRs embedded in a SMART with an ordinal outcome and discuss some combinatorial properties of this measure. Next, we propose a likelihood-based approach to estimate $GOR$ from SMART data, and derive the asymptotic properties of its estimate. We discuss  alternative ways to estimate $GOR$ using concordant-discordant pairs and two-sample $U$-statistic. We derive the required sample size formula for designing SMARTs with ordinal outcomes based on $GOR$. A simulation study shows the performance of the estimated $GOR$ in terms of the estimated power corresponding to the derived sample size. The methodology is applied to analyze data from the SMART+ study, conducted in the UK, to improve carbohydrate periodization behavior in athletes using a menu planner mobile application, Hexis Performance. A freely available Shiny web app using R is provided to make the proposed methodology accessible to other researchers and practitioners.}
\end{abstract}



\section{Introduction}\label{intro}
Personalized treatments constitute an increasingly important theme in today's health science. Dynamic treatment regimes (DTRs) \citep{murphy03, Chakraborty2014, Tsiatis2019} offer a vehicle to operationalize personalized interventions in time-varying settings. They are often used in the management of chronic conditions where a subject is typically treated at multiple stages, e.g., alcohol and drug abuse \citep{Ertefaie2016}, tobacco addiction \citep{chak09},  diabetes \citep{Luckett2019}, cancer \citep{Wang2012, Xu2016, Kidwell2018cancer}, HIV infection \citep{ROR2008}, and mental illnesses \citep{Zhang2018}. Precisely, DTRs are decision rules that recommend sequences of actions based on an individual's treatment and covariate history. Once constructed based on data, these rules can be employed to optimize the health outcome, depending on the individual's history.

\emph{Sequential Multiple Assignment Randomized Trial} (SMART) \citep{ld04, murphy05a, Almirall2014} is a special kind of clinical trial that provides high-quality data for comparing or constructing DTRs; the data from such trials are less vulnerable to causal confounding than longitudinal observational data. Methodological research on SMARTs has been on the rise in recent years, in accordance with the increasing prevalence of SMART or similar designs in practice, e.g., in cancer \citep{Wang2012}, chronic periodontitis \citep{xu2020smartp}, health behavior change  \citep{Almirall2014}, and substance abuse \citep{Dziak2019},  to mention a few. SMART designs involve randomization of individuals to available intervention options at an initial stage, followed by re-randomizations at each subsequent stage of some or all of the individuals to intervention options available at that stage. The re-randomizations and the set of intervention options at each stage may depend on information collected in prior stages such as how well the individual responded to the previous intervention(s). These designs attempt to conform better to the clinical practice, but still retain the well-known advantages of randomization over observational studies.

Various outcome types have been considered in the SMART design literature. For example, both \cite{ld04} and \citet{murphy05a} considered primary analysis of SMART design with continuous outcomes; details about related sample size calculations under a variety of research questions were given by \citet{oetting11}. These sample size calculations were further extended to cover binary outcomes by \cite{Ghosh2015}, noninferiority and equivalence SMART by \cite{Ghosh2020noninfi}, pilot SMART by \cite{yan2021sample}. Binary outcomes from a SMART were analyzed previously via likelihood-based methods by \citet{tms00} and recently a new simple data analytic approach was also proposed by \citet{Kidwell2018binary}. A lot of attention in the literature was focused on analysis of survival outcomes in a SMART \citep{kidwell2013} and associated sample size calculations \citep{Li_Murphy2011}. Composite outcomes were considered by \citet{Wang2012}. However, to the best of our knowledge, the literature on methods for analyzing ordinal outcomes in a SMART design till date is quite sparse, even though the outcomes of interest in many clinical and behavioral settings are measured in an ordinal scale and the methodologies to analyze such data in non-SMART context are available for decades \citep[e.g.,][]{Agresti1980}. The need for analysis methods to deal with ordinal outcomes in a SMART design has been acknowledged by \citet{Liu2014} for a long time. \citet{Zhao15a} proposed a classification-based optimization method that may be used to estimate optimal DTRs of ordinal outcomes, however the method was for analyzing observational data rather than SMART data. Our method is vastly different, whereby we employ an ordinal measure of association, the generalized odds ratio ($GOR$) \citep{Agresti1980}, that is interpretable and easy to use for practitioners. We will defer the explanation of $GOR$ to a later section. 


\subsection{SMART+ Study: Digital Approach to Improving Carbohydrate Periodization Behaviors in Athlete}\label{sec: smartplus}

\begin{figure}[ht]
	\centering
	\includegraphics[trim =4cm 17.5cm 4cm 0.5cm, height=9cm, width=10cm]{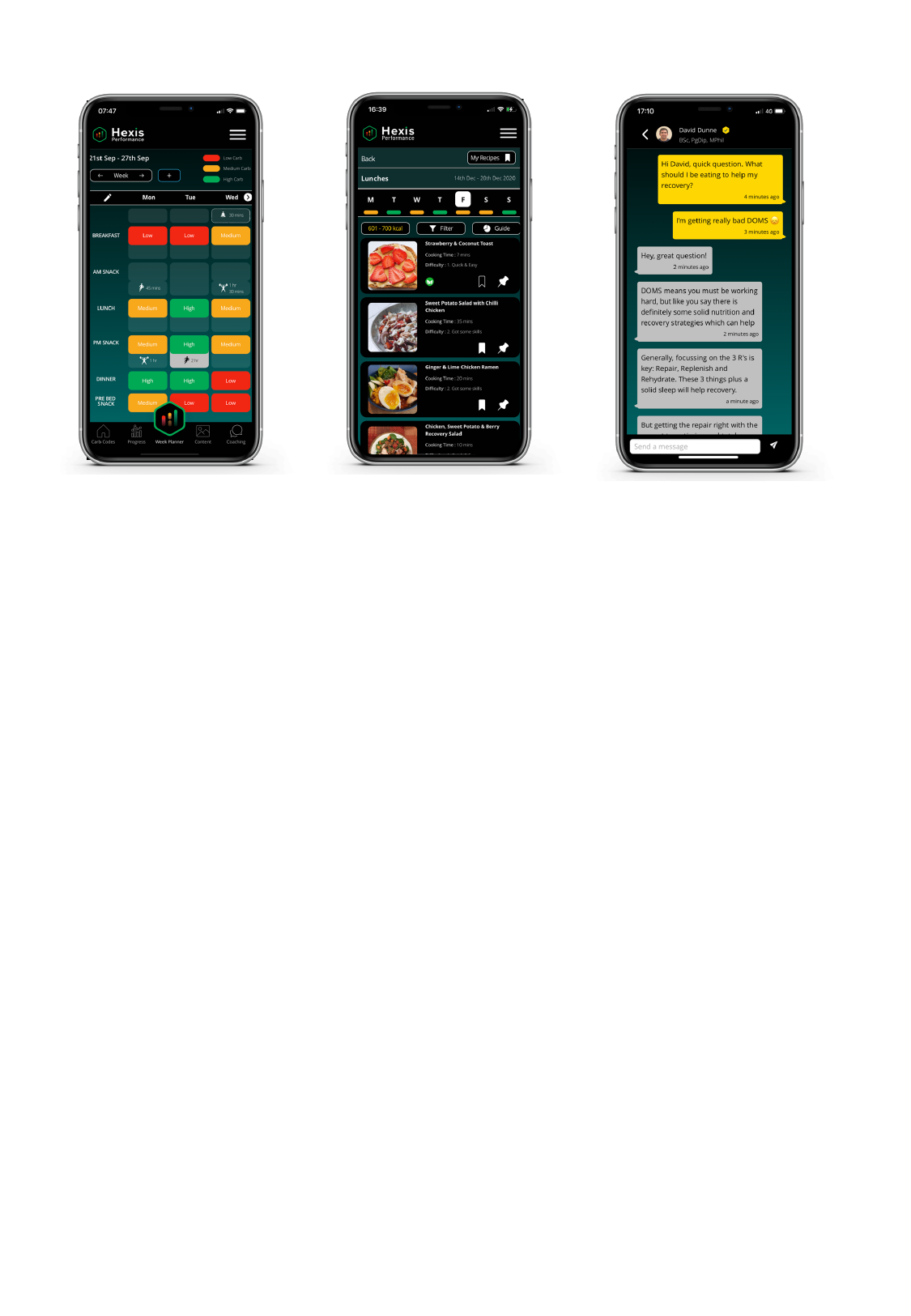}  
	\caption{Key screens of the Hexis Performance mobile application used in the SMART+ study. The messaging screen (right-most) is only activated when a nutrition coach (NC) is given. }
	\label{fig.0}
\end{figure}

\begin{figure}
	\centering
	\includegraphics[trim =5cm 0cm 0cm 0cm, height=11cm, width=12cm]{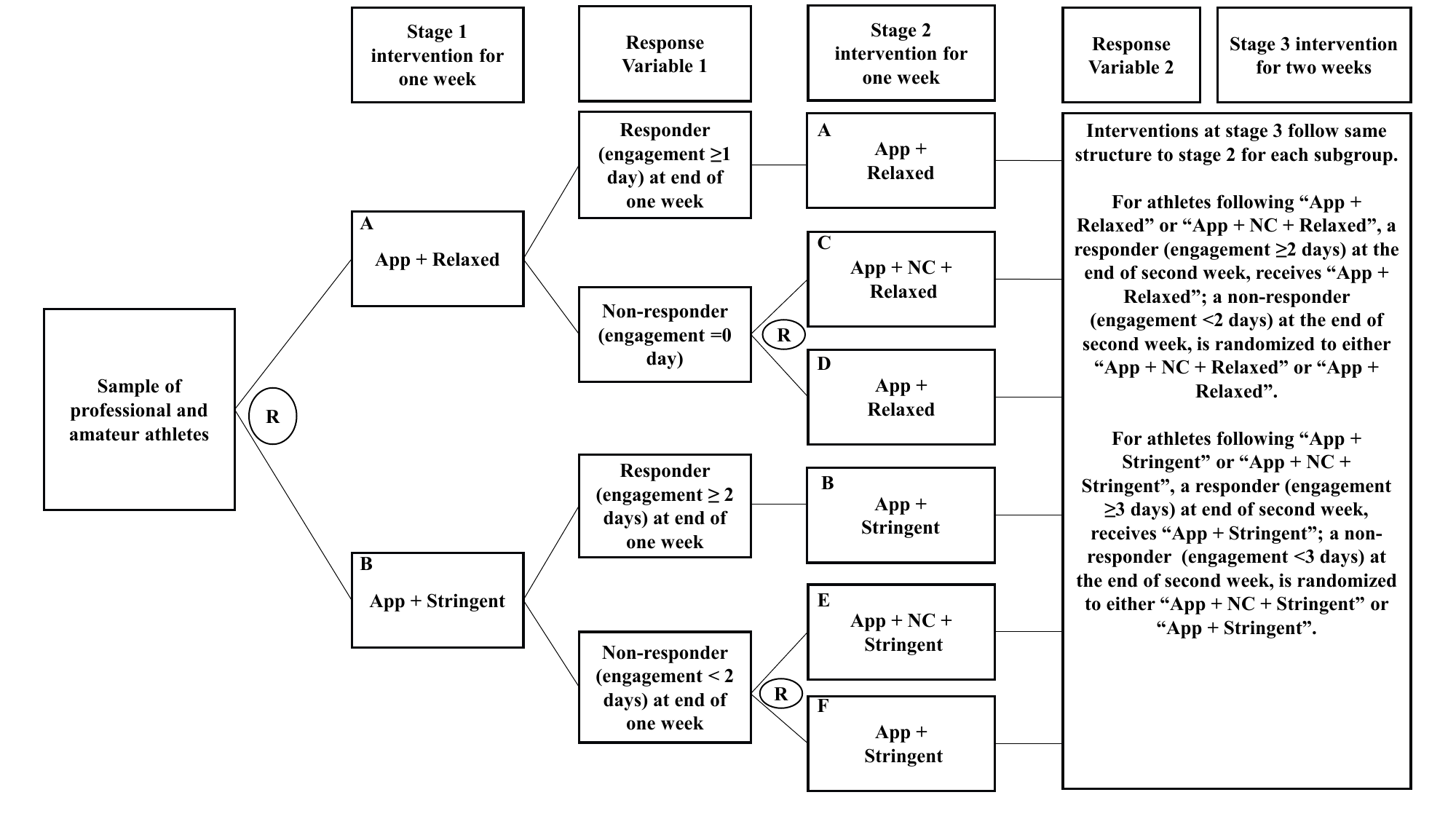}  
	\caption{The three-stage SMART design employed in the SMART+ study. The `R', in circle, represents randomization, `Relaxed' and `Stringent' represent the response criteria the athletes were assigned to follow throughout the trial. $A$ to $F$ are general terms to represent the interventions at stages 1 and 2.}
	\label{fig.1}
\end{figure}

Our work has been motivated by the digital approach to improving carbohydrate periodization behaviors in athlete (SMART+) study \citep{yan2022pilot} conducted in the UK, via a menu planner mobile application, {\em Hexis Performance} (Figure \ref{fig.0}). Carbohydrate periodization, also known as fueling for the work required, is the carbohydrate availability adjusted in accordance with the demands of the specific training session to be completed \citep{impey2018}, to optimize training performance. Despite the knowledge on what and when to eat, athletes usually struggle to adhere to the behavior, and often needs highly personalized support from nutrition coaches \citep{heikura2018}. As part of the study to reduce the need for expensive coaching support, a 4-week pilot three-stage SMART was conducted to evaluate the feasibility of developing a DTR around the {\em mobile application} (App), where the  {\em nutrition coach} (NC) communicates with the athletes only when necessary (at stage 2 and/or stage 3) depending on their engagement rates (response) on the App. The full SMART design used is shown in Figure \ref{fig.1}, wherein all athletes received the App and were randomized at stage 1 to either follow a relaxed or stringent response criteria for the rest of the trial. The full three-stage SMART design, with 16 embedded DTRs, is described by \citet{yan2022pilot}. 

The current study aims to assess feasibility, as such the primary outcomes focused on app and coaching uptake rates, and estimating the carbohydrate periodization success rates; there is less concern in formally comparing the clinical outcomes among the embedded regimes in the design. In this study, the clinical outcomes are carbohydrate periodization behavior ranking, carbohydrate periodization self-efficacy and belief about consequences, which are of ordinal nature, e.g., the carbohydrate periodization behavior ranking ranks athletes as 1 (not periodizing at all), 2 (periodizing energy/kcal only) or 3 (periodizing both energy/kcal and carbohydrates). If the study is to follow-up with a confirmatory trial, it is necessary to formally compare the regimes. However, existing methods are not tailored to compare any two embedded regimes in a SMART design with respect to ordinal outcomes. The methodologies developed in this article will help to investigate such research questions properly; see Section \ref{data_examp} for further details.

For simplicity of this article, we describe the trial as a two-stage design (the more conventional design) by ignoring the third stage, i.e. up to stage 2 in Figure \ref{fig.1}. An athlete assigned to the relaxed response criteria was defined as a responder after one week, if their engagement was for at least 1 day in that week, and a non-responder otherwise. For an athlete assigned to the stringent response criteria, the engagement had to be for at least 2 days in that week to be classified as a responder. Responders continued with the App, and non-responders were randomized to either continue with App alone or support with NC (App + NC) for the second week (stage 2). As such there are four embedded DTRs: 
$	d^{(1)} : (App + Relaxed, \{App + Relaxed\}^{R_r}\{App + NC + Relaxed\}^{1-R_r}),
d^{(2)} :  (App + Relaxed, \{App + Relaxed\}^{R_r}\{App + Relaxed\}^{1-R_r}),
d^{(3)} :(App + Stringent, \{App + Stringent\}^{R_s}\{App + NC + Stringent\}^{1-R_s}), \text{and }
d^{(4)} :(App + Stringent, \{App + Stringent\}^{R_s}\{App + Stringent\}^{1-R_s}),$
where $R_r$ and $R_s$ are the response indicators corresponding to the relaxed and stringent response criteria ( i.e., $R_{r} = 1$ and $R_{s} = 1$ if engagement rates are $\geq 1$ day and $\geq  2$ days), respectively. For example an athlete consistent with $d^{(1)}$ follows the regime ``give App to the athlete, if the athlete had no engagement with the App after one week, support App with NC in the second week, else continue with App".  It is worth noting that, although the athletes may receive the same intervention, the experience for individuals can be very different depending on their responses, e.g., a responder and a non-responder to stage 1 may both receive the App only at stage 2. Note that among the above four regimes, both $d^{(1)}$ and $d^{(2)}$ start with the same initial intervention $App + Relaxed$, and thus the responders at the first stage to $App + Relaxed$ are ``shared" between both these regimes. Following \citet{kidwell2013}, we call $d^{(1)}$ and $d^{(2)}$ {\em shared-path} DTRs. Similarly, $d^{(3)}$ and $d^{(4)}$ constitute another pair of shared-path DTRs. In contrast, $\{d^{(1)}, d^{(3)}\}$, $\{d^{(1)}, d^{(4)}\}$, $\{d^{(2)}, d^{(3)}\}$ and $\{d^{(2)}, d^{(4)}\}$ are pairs of distinct-path DTRs embedded in this design, as the elements within each pair start with different initial interventions and thus do not share any common group of trial participants.


\subsection{Generalized Odds Ratio}\label{sec: GOR}

A simple yet useful approach to compare ordinal outcomes across two or more groups utilizes a key quantity called the {\em generalized odds ratio} ($GOR$), first introduced by \citet{Agresti1980} and later employed in the context of standard two-group randomized controlled trials \citep[see, e.g.,][]{Lui2013}. As the name suggests, the $GOR$ is a generalization of the usual odds ratio for binary data, e.g., in case of contingency tables and logistic regression. Our key contribution in the current article is to generalize the notion of $GOR$ to more than one stages of the grouping variable (e.g., intervention) that can be applicable to a SMART design. 

The main difference between Agresti's $GOR$ and our newly proposed version is the presence of the response rate to the initial intervention within the definition of the odds ratio type quantity (note that these response rates can be different for different embedded regimes). We first present the general framework for $GOR$ using potential outcomes in Section \ref{sec: framework}.
Interestingly, the comparison between distinct-path regimes and that between shared-path regimes call for different considerations; hence we deal with these two cases separately in Sections \ref{Def_GOR_DP} and \ref{Def_GOR_SP}. In section \ref{Estimation-GOR}, we provide a likelihood based estimation of GOR, and a sample size calculation formula is given in  Section \ref{primary_ana}. We further present two alternatives to estimate $GOR$ in Section \ref{alt-est_GOR}, and highlight the broader scope of $GOR$, where it can also be computed for continuous outcomes using, e.g., $U$-statistics. The fact that $GOR$ is not restricted to ordinal outcomes opens up the possibility to compare DTRs beyond the standard methods based on mean outcomes or \textit{value functions} \citep[e.g.,][]{Zhao15b}. Section \ref{Simulation} shows the performance of the GOR using simulation studies. Finally, we apply our proposed methodology to the SMART+ data in Section \ref{data_examp} and conclude the article with a discussion in Section \ref{discussion}. To facilitate wide dissemination, we have also developed a Shiny web app implementing the methodology, which is freely available online (\textcolor{blue}{\url{https://www.iitg.ac.in/pgapps/dGOR/}}). In addition, in the Supplementary Materials, we extend the proposed methodology to a general $K$-stage SMART.

\section{A General Framework for Comparing DTRs with Ordinal Outcomes} \label{sec: framework}
Let $Y_{1}$ and $Y_{2}$ be two ordinal random variables, denoting the primary outcomes corresponding to two different groups (e.g., two intervention arms of a randomized controlled trial, RCT), each taking values in $J$ distinct ordered categories, say, $1,\ldots,J$. Then the generalized odds ratio ($GOR$) between groups 2 and 1 \citep{Agresti1980,Lui2013} is defined as $GOR_{(2,1)} = \frac{P(Y_{2}>Y_{1})}{P(Y_{2}<Y_{1})}$.
The interpretation of $GOR_{(2,1)}$ is simple. Assuming that a higher category of the outcome variable is better, $GOR_{(2,1)} >1$ indicates that the group 2 has a better outcome than group 1, and $GOR_{(2,1)} <1$ indicates the other way round; finally $GOR_{(2,1)}=1$ implies that there is no difference in outcomes between the two groups. In the following, we extend the {\em generalized odds ratio ($GOR$)} to allow us to compare DTRs embedded in a SMART with ordinal outcomes. 

To formally define the $GOR$ in the SMART context, we utilize the well-known \textit{potential outcomes} framework \citep{robins97}. For notational convenience, we re-represent the interventions in Figure \ref{fig.1} with more general terms, such that the embedded DTRs are rewritten as: $d^{(1)}: (A,A^{R_{A}}C^{1-R_{A}})$, $d^{(2)}: (A,A^{R_{A}}D^{1-R_{A}})$, $d^{(3)}: (B,B^{R_{B}}E^{1-R_{B}})$, and $d^{(4)}: (B,B^{R_{B}}F^{1-R_{B}})$, where $A, B, C, D, E$ and $F$ may be identical or different interventions, and $R_{A}$ and $R_{B}$ are the indicators of response (1 for responder, 0 for non-responder) corresponding to the initial interventions $A$ and $B$, respectively. Let $T_1$ and $T_2$ generically denote the interventions given to the subjects at the first and second stages, respectively, with $T_1 \in \{A, B\}$, $T_2 \in \{A, C, D\}$ if $T_1 = A$, and $T_2 \in \{B, E, F\}$ if $T_1 = B$. Then, let $Y_{T_1T_2}$ be the potential outcome under the intervention sequence $(T_1, T_2)$. Note that with respect to Figure \ref{fig.1} (up to two stages), there are only six potential outcomes, viz., $Y_{AA}, Y_{AC}, Y_{AD}, Y_{BB}, Y_{BE}$ and $Y_{BF}$. Then the potential outcome under any DTR in the current setup can be written in terms of the above six potential outcomes, as shown below. 

Consider the longitudinal data trajectory $(O_1, T_1, O_2, T_2, Y)$ corresponding to an individual subject participating in a SMART, where $O_k$ denotes the pre-intervention observations at stage $k\mbox{ } (k = 1,2)$, $T_k$ is the intervention given at stage $k\mbox{ } (k = 1,2)$ as defined before, and $Y$ is the primary outcome. Furthermore, define the history variables as $H_1 = O_1$ and $H_2 = (O_1, T_1, O_2)$. Note that the response indicator $R_{T_1}$ can be subsumed in $O_2$ ($R_{T_1}$ can be either a component or a low-dimensional summary of the vector-valued $O_2$). For example, in the SMART+ study,  $O_2$ is the number of days of engagement for that week, $R_{T_1}$ is an indicator for whether $O_2$ satisfies the relaxed or stringent response criteria following $T_1$ (i.e., at least 1 day and 2 days of engagement for athlete following $T_1 = App + Relaxed$ and $T_1 = App + Stringent$, respectively), and $Y$ is the carbohydrate periodization behavior ranking. 
Any arbitrary DTR $g$ with respect to the above data structure can be defined as a vector of decision rules, $g = (g_1, g_2)$, where $g_1(H_1) \in \mathcal{A}_1$ and $g_2(H_2) \in \mathcal{A}_2$, with $\mathcal{A}_k$ denoting the class of intervention options at stage $k\mbox{} (k = 1,2)$. Then the potential outcome under the arbitrary DTR $g$ can be defined as $
Y_g = Y_{AA} I \{g_1(H_1) = A, g_2(H_2) = A\} + Y_{AC} I \{g_1(H_1) = A, g_2(H_2) = C\}
+  Y_{AD} I \{g_1(H_1) = A, g_2(H_2) = D\} + Y_{BB} I \{g_1(H_1) = B, g_2(H_2) = B\}  
+  Y_{BE} I \{g_1(H_1) = B, g_2(H_2) = E\} + Y_{BF} I \{g_1(H_1) = B, g_2(H_2) = F\}, 
$ where $I\{\cdot\}$ is an indicator function. Now we are in a position to define $GOR$ as follows. 
Let $Y_{g}$ and $Y_{g'}$ denote the potential outcomes under two dynamic regimes $g$ and $g'$. Then the generalized odds ratio ($GOR$) between $g$ and $g'$, denoted $\eta_{g,g'}$, is defined as\useshortskip
\begin{align}
\eta_{g,g'} = \frac{P(Y_{g}>Y_{g'})}{P(Y_{g}<Y_{g'})},
\label{GOR_g}
\end{align}
where $P(Y_{g}>Y_{g'})$ is the probability that the outcome for a randomly selected subject from the set of subjects whose intervention sequence is consistent with the regime $g$ is larger than the outcome for a randomly selected subject from the set of subjects whose intervention sequence is consistent with the regime $g'$. 

\section{Comparison of Two Embedded Dynamic Regimes}\label{diff_ini_tret}
\subsection{Definition and properties of GOR for comparing distinct-path regimes}\label{Def_GOR_DP}
Suppose we are interested in comparing two distinct-path embedded dynamic regimes $d^{(1)}: (A,A^{R_{A}}C^{1-R_{A}})$ and $d^{(3)}: (B, B^{R_{B}}E^{1-R_{B}})$. This would be equivalent to $d^{(1)}:  (App + Relaxed,$ $ \{App + Relaxed\}^{R_r}\{App + NC + Relaxed\}^{1-R_r}) \mbox{ and }$
$d^{(3)}:  (App + Stringent, \{App + Stringent\}^{R_s} $ $\{App + NC + Stringent\}^{1-R_s})$ as previously described in Section \ref{intro}. In the current context, it turns out that $R_{A}= I \{d^{(1)}_1(H_1) = d^{(1)}_2(H_2) = A\}$ and $R_{B}= I \{d^{(3)}_1(H_1) = d^{(3)}_2(H_2) = B\}$. Also, let $Y_{d^{(1)}}$ denotes the primary outcome of a randomly selected subject from the set of subjects whose intervention sequence is consistent with the regime $d^{(1)}$. Similarly, $Y_{d^{(3)}}$ denotes the same corresponding to the regime $d^{(3)}$. Then,\useshortskip $P(Y_{d^{(3)}}>Y_{d^{(1)}}) $
\begin{tiny}
	\begin{eqnarray} 
	= \sum\limits_{R_{A}, R_{B} \in \{0, 1\}}\sum\limits_{u=1}^{J-1}\sum\limits_{s=u+1}^{J} \bigg \{ \gamma_{A}^{R_{A}} \gamma_{B}^{R_{B}}(1- \gamma_{A})^{1-R_{A}}(1- \gamma_{B})^{1-R_{B}}
	\pi_{AA^{R_{A}}C^{1-R_{A}},u}\pi_{BB^{R_{B}}E^{1-R_{B}},s} \bigg \}, \nonumber 
	\end{eqnarray}
\end{tiny}where $u$ corresponds to the response category of the regime $d^{(1)}$ and $s$ corresponds to the same for the regime $d^{(3)}$, $\pi_{AA,u} = P(Y_{AA} = u)$, $\pi_{AC,u} = P(Y_{AC} = u)$, $\pi_{BB,s} = P(Y_{BB} = s)$ and $\pi_{BE,s} = P(Y_{BE} = s)$.
Similarly $P(Y_{d^{(3)}}<Y_{d^{(1)}})$ can be computed. 
Hence, from (\ref{GOR_g}) the $GOR$ is given by $\eta_{d^{(3)},d^{(1)}}^{\text{DP}} $
\begin{tiny}
	\begin{eqnarray}
	= \frac{\sum\limits_{R_{A},R_{B} \in \{0, 1\}}\sum\limits_{u=1}^{J-1}\sum\limits_{s=u+1}^{J} \gamma_{A}^{R_{A}} \gamma_{B}^{R_{B}}(1- \gamma_{A})^{1-R_{A}}(1- \gamma_{B})^{1-R_{B}} \times \pi_{AA^{R_{A}}C^{1-R_{A}},u}\pi_{BB^{R_{B}}E^{1-R_{B}},s} }{\sum\limits_{R_{A}, R_{B} \in \{0, 1\}}\sum\limits_{u=2}^{J}\sum\limits_{s=1}^{u-1} \gamma_{A}^{R_{A}} \gamma_{B}^{R_{B}}(1- \gamma_{A})^{1-R_{A}}(1- \gamma_{B})^{1-R_{B}} \times \pi_{AA^{R_{A}}C^{1-R_{A}},u}\pi_{BB^{R_{B}}E^{1-R_{B}},s}}, 
	\label{GOR_expression}
	\end{eqnarray}
\end{tiny}where the superscript ``DP'' indicates that the two regimes under comparison are distinct-path regimes. Note that, for $J=2$, i.e. for binary outcome data, the above expression boils down to\useshortskip

\begin{tiny}
	\begin{eqnarray}
	\eta_{d^{(3)},d^{(1)}}^{\text{DP}} = \frac{\sum\limits_{R_{A}, R_{B} \in \{0, 1\}} \gamma_{A}^{R_{A}} \gamma_{B}^{R_{B}}(1- \gamma_{A})^{1-R_{A}}(1- \gamma_{B})^{1-R_{B}} \times \pi_{AA^{R_{A}}C^{1-R_{A}},1}\pi_{BB^{R_{B}}E^{1-R_{B}},2} }{\sum\limits_{R_{A}, R_{B} \in \{0, 1\}} \gamma_{A}^{R_{A}} \gamma_{B}^{R_{B}}(1- \gamma_{A})^{1-R_{A}}(1- \gamma_{B})^{1-R_{B}} \times \pi_{AA^{R_{A}}C^{1-R_{A}},2}\pi_{BB^{R_{B}}E^{1-R_{B}},1}},
	\label{OR_expression}
	\end{eqnarray} 
\end{tiny}which is simply the odds ratio ($OR$) for SMART data. The expression (\ref{OR_expression}) can further reduce to the conventional odds ratio, $\pi_{AC,1}\pi_{BE,2}/\pi_{AC,2}\pi_{BE,1}$, only when there is no split of trial subjects according to their response/non-response statuses at the end of stage 1; such a situation arises in a SMART involving smoking cessation interventions \citep{chak09}.


Note that, $\eta_{d^{(3)},d^{(1)}}^{\text{DP}}$ in (\ref{GOR_expression}) can be alternatively defined as\useshortskip

\begin{tiny}
	\begin{eqnarray}\label{GOR_matrix_exp_DP}
	\eta_{d^{(3)},d^{(1)}}^{\text{DP}} &=& \frac{\sum\limits_{R_{A}, R_{B} \in \{0, 1\}} \bigg\{ \gamma_{A}^{R_{A}} \gamma_{B}^{R_{B}}(1- \gamma_{A})^{1-R_{A}}(1- \gamma_{B})^{1-R_{B}} \biggr. }{\sum\limits_{R_{A}, R_{B} \in \{0, 1\}} \bigg\{ \gamma_{A}^{R_{A}} \gamma_{B}^{R_{B}}(1- \gamma_{A})^{1-R_{A}}(1- \gamma_{B})^{1-R_{B}}
		\biggr. }  \\ 
	&\times& \frac{ \biggl. [\mathbf{1'(U(\Pi_{AA^{R_{A}}C^{1-R_{A}}}\Pi_{BB^{R_{B}}E^{1-R_{B}}}^{'}) - diag(\Pi_{AA^{R_{A}}C^{1-R_{A}}}\Pi_{BB^{R_{B}}E^{1-R_{B}}}^{'})) 1} ] \bigg \}}{ \biggl. [\mathbf{1'(L(\Pi_{AA^{R_{A}}E^{1-R_{A}}}\Pi_{BB^{R_{B}}E^{1-R_{B}}}^{'}) - diag(\Pi_{AA^{R_{A}}C^{1-R_{A}}}\Pi_{BB^{R_{B}}E^{1-R_{B}}}^{'})) 1} ] \bigg \} } \nonumber
	\end{eqnarray}
\end{tiny}where the column vector $\Pi_{T_1T_1^{R_{T_1}}T_2^{1-R_{T_1}}}= (\pi_{T_1T_1^{R_{T_1}}T_2^{1-R_{T_1}},1},  \cdots, \pi_{T_1T_1^{R_{T_1}}T_2^{1-R_{T_1}},J})'$; $\mathbf{U(Z)}$, $\mathbf{L(Z)}$ and $\mathbf{diag(Z)}$ denote the upper triangular part, the lower triangular part and the diagonal part of a square matrix $\mathbf{Z}$ after replacing other elements by zeros, respectively. For example,
$$ Z = 
\begin{bmatrix}
a & b & c \\
d & e & f \\
g & h & i 
\end{bmatrix}
\mbox{implies } \mathbf{U(Z)} = 
\begin{bmatrix}
a & b & c \\
0 & e & f \\
0 & 0 & i 
\end{bmatrix},
\mathbf{L(Z)} = 
\begin{bmatrix}
a & 0 & 0 \\
d & e & 0 \\
g & h & i 
\end{bmatrix}
\mbox{and } \mathbf{diag(Z)} = 
\begin{bmatrix}
a & 0 & 0 \\
0 & e & 0 \\
0 & 0 & i 
\end{bmatrix}.
$$ 
The expression of $\eta_{d^{(3)},d^{(1)}}^{\text{DP}}$ in (\ref{GOR_matrix_exp_DP}) is computationally easier to work with. If we assume the cell probabilities of the ordinal outcome in the two responder arms (and the two  non-responder arms) to be same in two embedded distinct-path regimes, then the following proposition states the relationship between the $GOR$ and the two response probabilities corresponding to the two distinct-path regimes. 

\begin{proposition}\label{proposition_1}
	Let $\mathbf{\Pi_{AA} = \Pi_{BB}}$, $\mathbf{\Pi_{AC} = \Pi_{BE}}$ and $\mathbf{\Pi_{AA} \neq \Pi_{AC}}$, 
	\begin{enumerate}[topsep=0pt,itemsep=-1ex,partopsep=1ex,parsep=1ex]
		\item[i)] then $\eta_{d^{(3)},d^{(1)}}^{\text{DP}} = 1$ iff $\gamma_{A} = \gamma_{B}$,		
		\item[ii)] furthermore, let $\mathbf{1'(U(\Pi_{AA}\Pi_{BE}^{'})1 \gtrless 1'(L(\Pi_{AA}\Pi_{BE}^{'})1}$, then $\eta_{d^{(3)},d^{(1)}}^{\text{DP}} \gtrless 1$ iff $\gamma_{A} \gtrless \gamma_{B}$.
	\end{enumerate} 
\end{proposition}
Note that, $\mathbf{1'(U(\Pi_{AA}\Pi_{BE}^{'})1} = \sum\limits_{u=1}^{J}\sum\limits_{s=u}^{J} \pi_{AA,u} \pi_{BE,s}$ and $\mathbf{1'(L(\Pi_{AA}\Pi_{BE}^{'})1} = \sum\limits_{u=1}^{J}\sum\limits_{s=1}^{u}$ $ \pi_{AA,u} \pi_{BE,s}$.
In contrast with Proposition \ref{proposition_1}, if we assume the cell probabilities of the ordinal outcome in the responder arm of one regime are same as those of the non-responder arm of the other regime in two embedded distinct-path regimes, then the following proposition states the relationship between the $GOR$ and the two response probabilities corresponding to the two distinct-path regimes.
\begin{proposition}
	Let $\mathbf{\Pi_{AA} = \Pi_{BE}}$, $\mathbf{\Pi_{AC} = \Pi_{BB}}$ and $\mathbf{\Pi_{AA} \neq \Pi_{AC}}$, 
	\begin{enumerate}[topsep=0pt,itemsep=-1ex,partopsep=1ex,parsep=1ex]
		\item[i)] then $\eta_{d^{(3)},d^{(1)}}^{\text{DP}} = 1$ iff $\gamma_{A} = 1 - \gamma_{B}$,
		
		\item[ii)] furthermore, let $\mathbf{1'(U(\Pi_{AA}\Pi_{BB}^{'})1 \gtrless 1'(L(\Pi_{AA}\Pi_{BB}^{'})1}$, then $\eta_{d^{(3)},d^{(1)}}^{\text{DP}} \gtrless 1$ iff $\gamma_{A} \gtrless 1 - \gamma_{B}$. 
	\end{enumerate} 
\end{proposition}


\subsection{Definition and properties of GOR for comparing shared-path regimes}\label{Def_GOR_SP}
In this section, we are interested in comparing two shared-path dynamic regimes, using the same notations as in Section \ref{Def_GOR_DP}. Consider two embedded shared-path regimes $d^{(1)}: (A, A^{R_{A}}C^{1-R_{A}})$ and $d^{(2)}: (A, A^{R_{A}'}D^{1-R_{A}'})$. Here, $R_{A}= I \{d^{(1)}_1(H_1) = d^{(1)}_2(H_2) = A\}$ and $R_{A}' = I \{d^{(2)}_1(H_1) = d^{(2)}_2(H_2) = A\}$ denote the response indicators (1 for responder, 0 for non-responder) for the two randomly selected subjects from the regimes $d^{(1)}$ and $d^{(2)}$, respectively. A shared-path example from the SMART+ study is $d^{(1)}:  (App + Relaxed, \{App + Relaxed\}^{R_r}\{App + NC + Relaxed\}^{1-R_r}) \mbox{ and } $
$d^{(2)}:  (App + Relaxed, \{App + Relaxed\}^{R_r}\{App + Relaxed\}^{1-R_r}).$ Also, let $Y_{d^{(1)}}$ and $Y_{d^{(2)}}$ denote their primary outcomes. Now we have, 

\begin{tiny}
	\begin{eqnarray}
	&& P(Y_{d^{(2)}}>Y_{d^{(1)}}) = \sum\limits_{R_{A}, R_{A}' \in \{0, 1\}}\sum\limits_{u=1}^{J-1}\sum\limits_{s=u+1}^{J} \gamma_{A}^{R_{A} + R_{A}'} (1- \gamma_{A})^{2-R_{A} - R_{A}'} \times \pi_{AA^{R_{A}}C^{1-R_{A}},u}\pi_{AA^{R_{A}'}D^{1-R_{A}'},s}. \nonumber
	\end{eqnarray}
\end{tiny}Similarly $P(Y_{d^{(2)}}<Y_{d^{(1)}})$ can be computed. Hence, from (\ref{GOR_g}) the $GOR$ is given by\useshortskip
\begin{tiny}
	\begin{eqnarray}
	\eta_{d^{(2)},d^{(1)}}^{\text{SP}}  &=& \frac{ \sum\limits_{R_{A}, R_{A}' \in \{0, 1\}}\sum\limits_{u=1}^{J-1}\sum\limits_{s=u+1}^{J} \gamma_{A}^{R_{A} + R_{A}'} (1- \gamma_{A})^{2-R_{A} - R_{A}'} \times \pi_{AA^{R_{A}}C^{1-R_{A}},u}\pi_{AA^{R_{A}'}D^{1-R_{A}'},s} }{  \sum\limits_{R_{A}, R_{A}' \in \{0, 1\}}\sum\limits_{u=2}^{J}\sum\limits_{s=1}^{u-1} \gamma_{A}^{R_{A} + R_{A}'} (1- \gamma_{A})^{2-R_{A} - R_{A}'} \times \pi_{AA^{R_{A}}C^{1-R_{A}},u}\pi_{AA^{R_{A}'}D^{1-R_{A}'},s} }, \mbox{\hspace{4mm}}
	\label{GOR_expression_SP}
	\end{eqnarray}
\end{tiny}where the superscript ``SP'' indicates that the two regimes under comparison are shared-path regimes. The $GOR$, $\eta_{d^{(2)},d^{(1)}}^{\text{SP}}$ given in (\ref{GOR_expression_SP}) to compare two shared-path regimes is a special case of the $GOR$, $\eta_{d^{(3)},d^{(1)}}^{\text{DP}}$ defined in (\ref{GOR_expression}); $\eta_{d^{(2)},d^{(1)}}^{\text{SP}}$ can be obtained from the $\eta_{d^{(3)},d^{(1)}}^{\text{DP}}$ by replacing $d^{(3)}$, $\gamma_{B}$ and $R_{B}$ by $d^{(2)}$, $\gamma_{A}$ and $R_{A}'$, respectively. Note that $\eta_{d^{(2)},d^{(1)}}^{\text{SP}}$ in (\ref{GOR_expression_SP}) can be alternatively defined as\useshortskip
\begin{tiny}
	\begin{eqnarray}
	&& \eta_{d^{(2)},d^{(1)}}^{\text{SP}} \nonumber \\
	&=& \frac{\sum\limits_{R_{A}, R_{A}' \in \{0, 1\}}  \gamma_{A}(R_{A}, R_{A}') [\mathbf{1'(U(\Pi_{AA^{R_{A}}C^{1-R_{A}}}\Pi_{AA^{R_{A}'}D^{1-R_{A}'}}^{'}) - diag(\Pi_{AA^{R_{A}}C^{1-R_{A}}}\Pi_{AA^{R_{A}'}D^{1-R_{A}'}}^{'})) 1} ] }{\sum\limits_{R_{A}, R_{A}' \in \{0, 1\}} \gamma_{A}(R_{A}, R_{A}') [\mathbf{1'(L(\Pi_{AA^{R_{A}}C^{1-R_{A}}}\Pi_{AA^{R_{A}'}D^{1-R_{A}'}}^{'}) - diag(\Pi_{AA^{R_{A}}C^{1-R_{A}}}\Pi_{AA^{R_{A}'}D^{1-R_{A}'}}^{'})) 1} ]}, \nonumber
	\end{eqnarray}
\end{tiny}where $\gamma_{A}(R_{A}, R_{A}') = \gamma_{A}^{R_{A} + R_{A}'} (1- \gamma_{A})^{2-R_{A} - R_{A}'}$.


Note that, $\mathbf{\Pi_{AC} = \Pi_{AD}}$ implies $\eta_{d^{(2)},d^{(1)}}^{\text{SP}} = 1$. However, the converse is not true. We illustrate this point with the following example. Let $\Pi_{AA} = (0.2, 0.3, 0.5)$, $\Pi_{AC} = (0.12, 0.32, 0.56)$, $\Pi_{AD} = (0.06, 0.41, 0.53)$ and $\gamma_{A} =0.2$, here $GOR = 1$ even though $\mathbf{\Pi_{AC} \neq \Pi_{AD}}$. Also, $\eta_{d^{(2)},d^{(1)}}^{\text{SP}}$ is not invariant in $\mathbf{\Pi_{AA}}$. Consider two values of $\mathbf{\Pi_{AA}}$ as $(0.5, 0.4, 0.1)$ and $(0.2, 0.4, 0.4)$. It can be shown that for $\gamma_A = 0.2$, $\eta_{d^{(2)},d^{(1)}}^{\text{SP}}$ values are different in two cases (0.43 and 0.45) when both $\mathbf{\Pi_{AC}}$ and $\mathbf{\Pi_{AD}}$ are fixed at (0.3, 0.3, 0.4) and (0.6, 0.2, 0.2), respectively. 

\section{Maximum Likelihood Estimation of GOR}\label{Estimation-GOR}
We assume that the usual assumptions about potential outcomes \citep{robins97, robins04} in a longitudinal setting, viz., (i) \textit{consistency}, (ii) \textit{no unmeasured confounding} and (iii) \textit{positivity}, hold. Specifically, the consistency assumption states that the potential outcome under the observed intervention sequence and the observed outcome agree, i.e., $Y = Y_{T_1T_2}$ if the observed intervention sequence is indeed $(T_1, T_2)$. Thus, while the $GOR$ is defined conceptually using potential outcomes, it can be computed based on observed data. On the other hand, the no unmeasured confounding assumption states that intervention allocation is independent of future potential outcomes given the history; this is satisfied by design in case of SMARTs \citep{murphy05a}. The positivity assumption states that an individual has a positive probability of receiving any of the intervention sequences considered in the study. Thus, by positivity assumption, at least some individuals receive each of the possible intervention sequences $(T_1, T_2)$ feasible in the study.

The maximum likelihood estimate of $\eta_{d^{(3)},d^{(1)}}^{\text{DP}}$, say $\hat\eta_{d^{(3)},d^{(1)}}^{\text{DP}}$, can be computed by plugging-in the maximum likelihood estimates of $\gamma$s (empirical response rates) and $\pi$s (empirical probabilities of outcome categories) obtained from the data likelihood (see the Supplementary Materials,  Section 3). 
The asymptotic distribution of estimated $GOR$ in (\ref{GOR_expression}) is given by\useshortskip
\begin{eqnarray}
\sqrt{N}\Big(\hat\eta_{d^{(3)},d^{(1)}}^{\text{DP}} - \eta_{d^{(3)},d^{(1)}}^{\text{DP}}\Big) \rightarrow Normal(0, \sigma_{d^{(3)},d^{(1)}}^{2}), 
\label{asym_dist_DP}
\end{eqnarray}
where $N^{-1}\sigma_{d^{(3)},d^{(1)}}^{2}$ is the variance of $\hat\eta_{d^{(3)},d^{(1)}}^{\text{DP}}$, with $N$ being the total number of subjects in the trial. See the the Supplementary Materials for detailed derivation. Using delta method, we have\useshortskip
\begin{eqnarray}
\sqrt{N}\Big(\log \hat\eta_{d^{(3)},d^{(1)}}^{\text{DP}} - \log \eta_{d^{(3)},d^{(1)}}^{\text{DP}}\Big) \rightarrow Normal(0, \sigma_{d^{(3)},d^{(1)}}^{2}/(\eta_{d^{(3)},d^{(1)}}^{\text{DP}})^2). 
\label{Log_dist_DP}
\end{eqnarray}
For small $N$, use of (\ref{Log_dist_DP}) is preferable.
The maximum likelihood estimate of $\eta_{d^{(2)},d^{(1)}}^{\text{SP}}$, say $\hat\eta_{d^{(2)},d^{(1)}}^{\text{SP}}$, can be computed by plugging-in the maximum likelihood estimates of $\gamma$s and $\pi$s, as before. The asymptotic distribution is given by\useshortskip
\begin{eqnarray}
\sqrt{N}\Big(\hat\eta_{d^{(2)},d^{(1)}}^{\text{SP}} - \eta_{d^{(2)},d^{(1)}}^{\text{SP}}\Big) \rightarrow Normal(0, \sigma_{d^{(2)},d^{(1)}}^{2}), 
\label{asym_dist_SP}
\end{eqnarray}
where $N^{-1}\sigma_{d^{(2)},d^{(1)}}^{2}$ is the variance of $\hat\eta_{d^{(2)},d^{(1)}}^{\text{SP}}$. See the the Supplementary Materials for detailed derivation. Using delta method, we have\useshortskip 
\begin{eqnarray}
\sqrt{N}\Big(\log \hat\eta_{d^{(2)},d^{(1)}}^{\text{SP}} - \log \eta_{d^{(2)},d^{(1)}}^{\text{SP}}\Big) \rightarrow Normal(0, \sigma_{d^{(2)},d^{(1)}}^{2}/(\eta_{d^{(2)},d^{(1)}}^{\text{SP}})^2).  
\label{Log_dist_SP}
\end{eqnarray}

\subsection{Primary analysis and sample size formula}\label{primary_ana}
Specifying a primary analysis is necessary for powering a SMART \citep{murphy05a}. If the current pilot SMART+ study were to proceed as a full-fledged SMART, a reasonable primary analysis would be to test if the regime $d^{(3)}$ differs from the regime $d^{(1)}$ in terms of the ordinal primary outcome (comparison of distinct-path embedded regimes). Consider the null hypothesis $H_0: \eta_{d^{(3)},d^{(1)}}^{\text{DP}}=1$ against the alternative hypothesis $H_1: \eta_{d^{(3)},d^{(1)}}^{\text{DP}} = e^{\delta}$, where $\delta$ could take any positive or negative value. Using (\ref{asym_dist_DP}), we use $\hat\eta_{d^{(3)},d^{(1)}}^{\text{DP}}$ as the test statistic for the primary analysis. For a positive (negative) value of $\delta$, the high (low) value of $\hat\eta_{d^{(3)},d^{(1)}}^{\text{DP}}$ is an indication of departure from the null hypothesis. Let $z_{\alpha/2}$ be the $(1-\alpha/2)$ percentile of the standard normal distribution and set the power of the test as $1- \beta$, where $\beta$ is the type-II error. Using (\ref{asym_dist_DP}) the required sample size is given by $N = ( z_{\alpha/2} + z_{\beta})^2 \frac{\sigma_{d^{(3)},d^{(1)}}^{2}}{\delta^2}. $
Here $N$ is the total number of subjects in the trial. We can consider $\delta/\sigma_{d^{(3)},d^{(1)}}$ as the standardized effect size, which can potentially be elucidated from scientific investigators prior to designing the SMART. The primary analysis and the sample size formula for the shared-path setup are similar to the above with replacement of $\eta_{d^{(3)},d^{(1)}}^{\text{DP}}$ and $\sigma_{d^{(3)},d^{(1)}}^{2}$ by $\eta_{d^{(2)},d^{(1)}}^{\text{SP}}$ and $\sigma_{d^{(2)},d^{(1)}}^{2}$, respectively.

\section{Alternative Ways to Estimate GOR}\label{alt-est_GOR}
\subsection{GOR based on Concordant and Discordant pairs}
\cite{Goodman1954} proposed a measure of association $\Gamma$ for an $I \times J$ cross-classified table, where $I$ and $J$ represent the number of ordinal categories of the row variable and the column variable, respectively. The Goodman-Kruskal $\Gamma$ is based on the number of concordant and discordant pairs corresponding to any two individuals randomly chosen from the population. Let the two chosen individuals be denoted by $(i,j)$ and $(i', j')$, where $i,i' = 1, \cdots, I$ and $j,j' = 1,\cdots, J$. The chosen pair is called concordant if ($i < i'$ and $j < j'$) or ($i > i'$ and $j > j'$) and discordant if ($i < i'$ and $j > j'$) or ($i > i'$ and $j < j'$). Note that, the $GOR$ described in the current article is not defined for a $I \times J$ cross-classified table; rather, we restrict to a $2 \times J$ table where the two rows denote two sub-populations (e.g., two arms of an RCT) and the columns correspond to the $J$ ordered categories of an ordinal variable $Y$. We can then calculate $GOR$ based on the number of concordant and discordant pairs in a slightly different manner. Suppose a randomly selected individual from the sub-population 1 is denoted by $(1,u)$ and that from the sub-population 2 is denoted by $(2, s)$; $u,s = 1,\cdots,J$. Define the randomly selected pair  to be concordant if $u<s$, i.e., the individual selected from the sub-population 2 has higher response category than the individual selected from the sub-population 1, and discordant if $u >s$. Thus, the total number of concordant and discordant pairs are given by $\sum_{u=1}^{J-1}\sum_{s=u+1}^{J}n_{1u}n_{2s}$ and $\sum_{u=2}^{J}\sum_{s=1}^{u-1}n_{1u}n_{2s}$, respectively, where $n_{ij}$ denotes the cell frequency corresponding to the $i^{th}$ sub-population and the $j^{th}$ ordinal category with $i=1,2; j=1,\cdots,J$. One can estimate $P(Y_{2}>Y_{1})$ by $\frac{1}{n_{1\cdot}n_{2\cdot}}\sum_{u=1}^{J-1}\sum_{s=u+1}^{J}n_{1u}n_{2s}$, where $n_{i\cdot} = \sum_{u=1}^{J}n_{iu}$, $i=1,2$ and thus can estimate $GOR_{(2,1)}$ as the ratio of the total number of concordant pairs over the total number of discordant pairs as $\widehat{GOR}_{(2,1)} = \frac{\sum_{u=1}^{J-1}\sum_{s=u+1}^{J}n_{1u}n_{2s}}{\sum_{u=2}^{J}\sum_{s=1}^{u-1}n_{1u}n_{2s}}. $

Similarly, for a SMART design, we can express the estimate of $GOR$ defined in (\ref{GOR_expression}) as\useshortskip

\begin{tiny}
	\begin{eqnarray}
	\hat\eta_{d^{(3)},d^{(1)}}^{\text{DP}}  = \frac{\sum\limits_{u=1}^{J-1}\sum\limits_{s=u+1}^{J}\bigg[ n_{AA,u}n_{BB,s} + 2 \times n_{AA,u}n_{BE,s} + 2 \times n_{AC,u}n_{BB,s} + 4 \times n_{AC,u}n_{BE,s}  \bigg]}{\sum\limits_{u=2}^{J}\sum\limits_{s=1}^{u-1}\bigg[  n_{AA,u}n_{BB,s} + 2 \times n_{AA,u}n_{BE,s} + 2 \times n_{AC,u}n_{BB,s} + 4 \times n_{AC,u}n_{BE,s} \bigg]}, \hspace{0.25cm}
	\label{GOR_DP_concord}
	\end{eqnarray}
\end{tiny}where $n_{AA,u}$ denotes the cell frequency of the $u^{th}$ ordinal category in the responder arm of the regime $d^{(1)}$; other cell frequencies are defined accordingly. Note that, the total number of individuals in the responder branch of the regime $d^{(1)}$ is $n_{AA\cdot} = \sum_{u=1}^{J} n_{AA,u} = \frac{N}{2}\times \hat\gamma_{A}$, where $N$ is the known total number of individuals in the entire SMART. In (\ref{GOR_DP_concord}), the four components in the numerator (or the denominator) refer to the four different ways of choosing a pair of individuals from the two regimes $d^{(1)}$ and $d^{(3)}$. Specifically, we have the set as $\{(AA, BB), (AA, BE), (AC, BB), (AC, BE)\}$, where $(AC, BB)$ refers to a pair wherein the first individual is randomly chosen from the non-responder branch of the regime $d^{(1)}$ and the second individual is randomly chosen from the responder branch of the regime $d^{(3)}$. Intuitively, $\hat\eta_{d^{(3)},d^{(1)}}^{\text{DP}}$ is the ratio of the weighted sum of the concordances to the weighted sum of the discordances, where the weights are the inverse probability weights associated with a ``restricted" SMART as in Figure \ref{fig.1} \citep{robins97, murphy05a, Nahum-Shani2012a}. More specifically, the weight is a product of the number of times each of the selected two individuals in the pair is (evenly) randomized. For example, if the pair is coming from $(AC, BB)$, the weight is $2 \times 1$ because the individual from the regime $d^{(1)}$ is a non-responder and hence randomized twice, whereas the one from the regime $d^{(3)}$ is a responder and hence randomized only once. However, for the ``unrestricted" SMART design \citep{Nahum2019_education} where both responders and non-responders are evenly randomized (See Supplementary Material Figure 1), no such weighting is necessary. The use of the notion of concordance in the DTR literature is not entirely new; see \cite{Fan2017} for concordance-assisted learning for optimal treatment regimes.

\subsection{GOR for continuous outcomes based on U-statistic}\label{sec: u-stat}
The main focus of this article is to define and use $GOR$ for comparing DTRs with ordinal outcomes. However, the $GOR$ defined in (\ref{GOR_g}) does not require $Y_{g}$ and $Y_{g'}$ to be ordinal variables. In fact, as discussed by \citet{Agresti1980} in the context of two-grouped $GOR$, the $GOR$ can as well be used to compare DTRs based on continuous outcomes. Furthermore, the $GOR$ defined in (\ref{GOR_g}) does not assume any particular parametric model for the data. Thus the scope of $GOR$ is very broad, and it provides an alternative approach to comparing DTRs beyond the standard methods based on mean potential outcomes or {\em value functions} \citep{zhao12, Zhao15b}; the value function of a regime $g$ is defined as $E(Y_g)$. Thus, while value-based methods work with mean potential outcomes, $GOR$-based methods work with stochastic ordering of potential outcomes. 

We can write the $GOR$ defined in (\ref{GOR_g}) to compare the regimes $d^{(3)}$ and $d^{(1)}$ in Figure \ref{fig.1} with continuous outcome as $ \eta_{d^{(3)},d^{(1)}} = \frac{P(Y_{d^{(3)}}>Y_{d^{(1)}})}{1-P(Y_{d^{(3)}}>Y_{d^{(1)}})}$.
Note that, the $P(Y_{g}>Y_{g'})$ can be estimated by using the $U$-statistic\useshortskip
\begin{eqnarray}
\frac{1}{n_{d^{(3)}}n_{d^{(1)}}} \sum\limits_{s=1}^{n_{d^{(3)}}} \sum\limits_{u=1}^{n_{d^{(1)}}} \phi(Y_{d^{(3)} s}; Y_{d^{(1)} u}),
\label{u-stat}
\end{eqnarray}
where $n_{d^{(3)}}$ and $n_{d^{(1)}}$ denote the total number of individuals with intervention sequences consistent with the regimes $d^{(3)}$ and $d^{(1)}$, respectively; $Y_{d^{(3)} s}$ denotes the outcome of $s^{th}$ individual corresponding to the regime $d^{(3)}$;  and\useshortskip
\begin{tiny}
	\begin{eqnarray}
	\phi(Y_{d^{(3)} s}; Y_{d^{(1)} u}) = \sum\limits_{R_B, R_A \in \{0,1\}} \gamma_{A}^{R_{A}} \gamma_{B}^{R_{B}}(1- \gamma_{A})^{1-R_{A}}(1- \gamma_{B})^{1-R_{B}} 
	I\{Y_{BB^{R_{B}}E^{1-R_{B}},s} > Y_{AA^{R_{A}}C^{1-R_{A}},u}\}, \nonumber
	\end{eqnarray}
\end{tiny}
where $I\{\cdot\}$ is an indicator function; $Y_{BE,s}$ denotes the outcome of $s^{th}$ individual with responder status $R_{B} = 0$  (i.e, $(T_1, T_2) = (B, E)$) corresponding to the regime $d^{(3)}$. The $U$-statistic in (\ref{u-stat}) is a two-sample $U$-statistic with $E(\phi(Y_{d^{(3)} s}; Y_{d^{(1)} u})) = P(Y_{d^{(3)}}>Y_{d^{(1)}})$ and $\phi(\cdot; \cdot)$ is symmetric by default as it contains only one argument from each of the two samples \citep{Lehmann_large_sample}.

\section{Simulation Studies}\label{Simulation}
Before we apply the proposed methodologies to the SMART+ data, we present two thorough simulation studies to illustrate their performances. Specifically, Section \ref{simu_DP} and \ref{simu_SP} present the performance of the $GOR$-based estimation and inference in case of distinct-path and shared-path dynamic regimes, respectively (cf. Section \ref{diff_ini_tret}); the details of the data generation process are described in the \textit{Supplementary Materials}. The response rates corresponding to initial interventions $A$ and $B$ are taken as $\gamma_A = 0.3$ and $\gamma_B = 0.4$, respectively. Simulations considering different values of the response rates are shown in the \textit{Supplementary Materials}. We set the type-I error rate at 0.05 and the nominal power at 0.80. Section \ref{simu_small_prob} presents the results when cell probabilities are small. 

\subsection{Simulation study for distinct-path regimes}\label{simu_DP}
\begin{figure}[h]
	\centering
	\subfloat[ ]{{\includegraphics[trim =2cm 1cm 4cm 0cm, width=6cm]{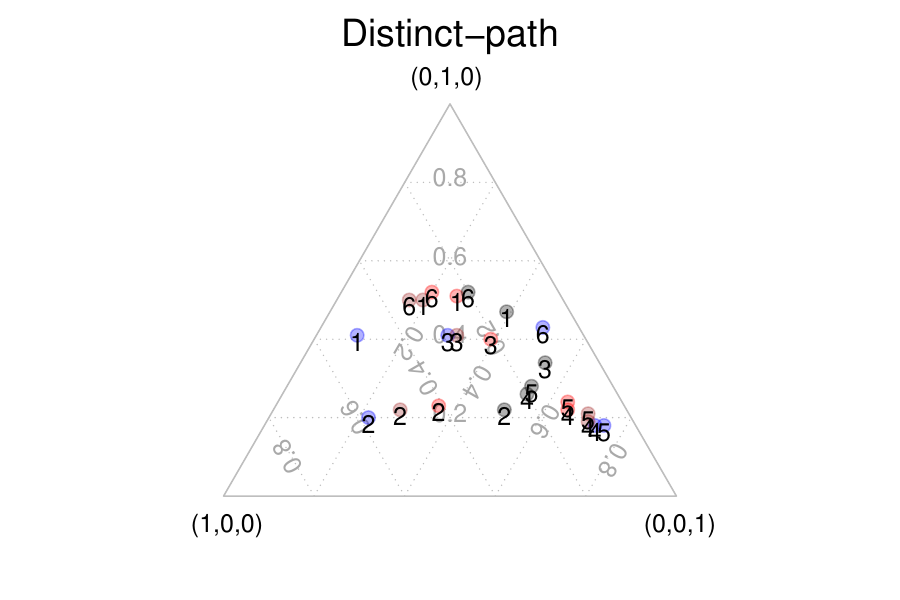} }}%
	\qquad
	\subfloat[ ]{{\includegraphics[trim = 2cm 1cm 4cm 0cm, width=6cm]{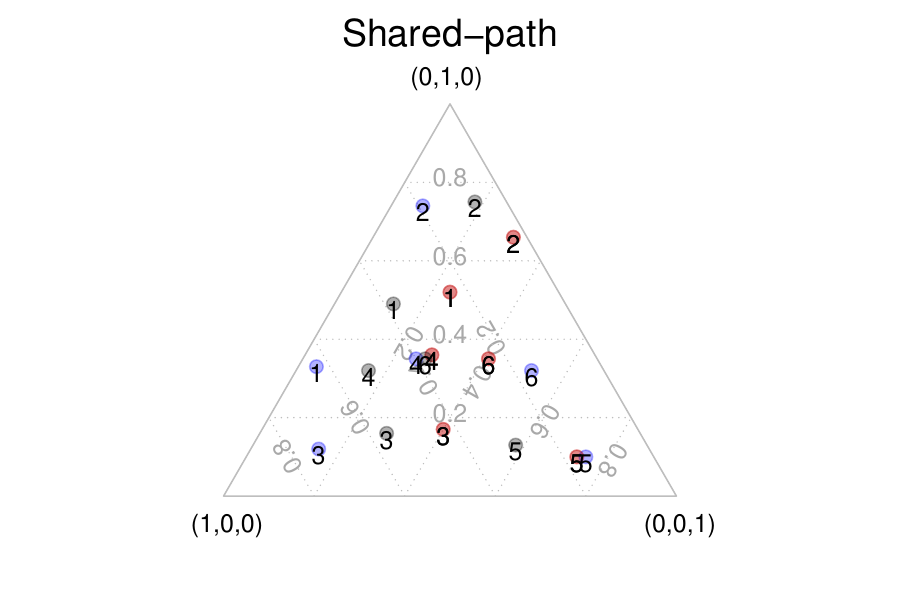} }}%
	\caption{Corresponding to Tables \ref{Simu_tab_DP} and \ref{Simu_tab_SP}, the Barycentric plots of different multinomial probabilities for comparing (a) distinct-path regimes,  \textcolor{red!40}{$\boldsymbol{\bullet}$}: $\pi(A, A)$ , \textcolor{blue!40}{$\boldsymbol{\bullet}$}: $\pi(A, C)$, \textcolor{Sepia!30}{$\boldsymbol{\bullet}$}: $\pi(B, B)$, \textcolor{black!40}{$\boldsymbol{\bullet}$}: $\pi(B, E)$;
		and (b) shared-path regimes, \textcolor{red!40}{$\boldsymbol{\bullet}$}: $\pi(A, A)$ , \textcolor{blue!40}{$\boldsymbol{\bullet}$}: $\pi(A, C)$,  \textcolor{black!40}{$\boldsymbol{\bullet}$}: $\pi(A, D)$. }
	\label{Bari_plot}%
\end{figure}

This section aims to assess the performance of $\eta_{d^{(3)},d^{(1)}}^{\text{DP}}$ in comparing two distinct-path embedded regimes $d^{(1)}$ and $d^{(3)}$ in a two-stage SMART. Taking the carbohydrate periodization behavior ranking in the SMART+ study for example, we consider the ordinal outcome $Y$ with $J=3$ categories in ascending order. For regime $d^{(1)}$, the cell probabilities of $Y$ are given by $\pi(A, A) = (\pi_{AA,1}, \pi_{AA,2}, 1-\pi_{AA,1} - \pi_{AA,2})$ for the responders and $\pi(A, C) = (\pi_{AC,1}, \pi_{AC,2}, 1-\pi_{AC,1} - \pi_{AC,2})$ for the non-responders. Likewise, for regime $d^{(3)}$, the corresponding cell probabilities are $\pi(B, B) = (\pi_{BB,1}, \pi_{BB,2}, 1-\pi_{BB,1} - \pi_{BB,2})$ for the responders and $\pi(B, E) = (\pi_{BE,1}, \pi_{BE,2}, 1-\pi_{BE,1} - \pi_{BE,2})$ for the non-responders. The serial number in the first column of Table \ref{Simu_tab_DP} shows six different scenarios. The first three of them correspond to $\eta_{d^{(3)},d^{(1)}}^{\text{DP}} > 1$ and the last three correspond to $\eta_{d^{(3)},d^{(1)}}^{\text{DP}} < 1$. Here $\eta_{d^{(3)},d^{(1)}}^{\text{DP}}$ denotes the true values of the corresponding population $GOR$. We obtain the true value of $GOR$ ($\eta_{d^{(3)},d^{(1)}}^{\text{DP}}$) by using Monte Carlo computation considering a `large' population of size of $10^{6}$. Figure \ref{Bari_plot}(a) shows the plot of \textit{barycentric coordinates} \citep{Jupp_tplot_packageR} of the probabilities from the six different scenarios in Table \ref{Simu_tab_DP}. Any three probabilities of the form $(p_1, p_2, 1 - p_1 - p_2)$ can be represented by a unique point in an equilateral triangle in a barycentric coordinate system. Three vertices are denoted by $(1, 0, 0), (0,1,0)$ and $(0, 0, 1)$. In a barycentric coordinate plot, a point $(p_1, p_2, 1 - p_1 - p_2)$ can be located by considering (i) a distance $p_1$ from the opposite arm of the vertex $(1,0,0)$, (ii) a distance $p_2$ from the opposite arm of the vertex $(0,1,0)$, and, (iii) a distance $1-p_1-p_2$ from the opposite arm of the vertex $(0,0,1)$. The objective of the Figure \ref{Bari_plot}(a) is to show how the six simulation scenarios in Table \ref{Simu_tab_DP} are distributed in the parameter space represented by the barycentric coordinate system.


In Table \ref{Simu_tab_DP}, the `Std.ES' denotes the standardized effect size which is calculated as $\log\eta_{d^{(3)},d^{(1)}}^{\text{DP}}$ divided by the square root of the variance of $\log\eta_{d^{(3)},d^{(1)}}^{\text{DP}}$; $N$ denotes the estimated sample size using the distribution of $\log\eta_{d^{(3)},d^{(1)}}^{\text{DP}}$. Based on the 5,000 simulations, $\hat\eta_{d^{(3)},d^{(1)}}^{\text{DP}}$ is the estimated value of $\eta_{d^{(3)},d^{(1)}}^{\text{DP}}$; SSE is the sample standard error; ASE is the asymptotic standard error; $\widehat{power}$ is the estimated power and $\widehat{CP}$ is the estimated coverage probability. In Table \ref{Simu_tab_DP}, $\hat\eta_{d^{(3)},d^{(1)}}^{\text{DP}}$ is close to its true value in all the six scenarios, indicating good estimation. The estimated sample size $N$ varies from 164 to 571 when $\eta_{d^{(3)},d^{(1)}}^{\text{DP}} >1$ and from 305 to 1096 when $\eta_{d^{(3)},d^{(1)}}^{\text{DP}} < 1$. Note that, $N$ is a decreasing function of the absolute value of the standardized effect size. However, we can have the same standardized effect size for different combinations of $\eta_{d^{(3)},d^{(1)}}^{\text{DP}}$, the response rates and the cell probabilities corresponding to the two regimes. The SSE and ASE are close to each other in all the scenarios. The estimated empirical powers and the coverage probabilities are close to their corresponding nominal values in all the cases.

\subsection{Simulation study for shared-path regimes}\label{simu_SP}
Here, we consider a simulation study to compare the shared-path regimes $d^{(1)}$ and $d^{(2)}$ that start with the same initial interventions as described in Section \ref{intro} and \ref{Def_GOR_SP}. As in the previous section, we set the number of categories of the ordinal outcome $Y$ as  $J=3$ in ascending order. For regime $d^{(2)}$, the cell probabilities of $Y$ are given by $\pi(A, A) = (\pi_{AA,1}, \pi_{AA,2}, 1-\pi_{AA,1} - \pi_{AA,2})$ for the responders and $\pi(A, D) = (\pi_{AD,1}, \pi_{AD,2}, 1-\pi_{AD,1} - \pi_{AD,2})$ for the non-responders. Note that, $\pi(A, A)$ corresponds to both regimes $d^{(1)}$ and $d^{(2)}$ (shared-path). The barycentric coordinate plot in Figure \ref{Bari_plot}(b) shows how the six simulation scenarios in Table \ref{Simu_tab_SP} are distributed in the barycentric coordinate system. 

Table \ref{Simu_tab_SP} shows the results of six scenarios for shared-path comparison of two regimes. As in Section \ref{simu_DP}, the first three scenarios correspond to $\eta_{d^{(2)},d^{(1)}}^{\text{SP}} > 1$ and the last three scenarios to $\eta_{d^{(2)},d^{(1)}}^{\text{SP}} < 1$. The estimated sample size $N$ ranges from 304 to 659 when $\eta_{d^{(2)},d^{(1)}}^{\text{SP}} > 1$ and from 241 to 1218 when  $\eta_{d^{(2)},d^{(1)}}^{\text{SP}} < 1$. In all the six scenarios, the SSE and the ASE are close to each other. The estimated empirical powers and the coverage probabilities are close to their corresponding nominal values.

In summary, from both the simulation studies in Sections \ref{simu_DP} and \ref{simu_SP}, it is evident that the proposed $GOR$s for comparing both distinct-path and shared-path regimes perform well.

\subsection{Simulation study with smaller cell probabilities}\label{simu_small_prob}

\begin{figure}%
	\centering
	\subfloat[ ]{{\includegraphics[trim =2cm 1cm 4cm 0cm, width=6cm]{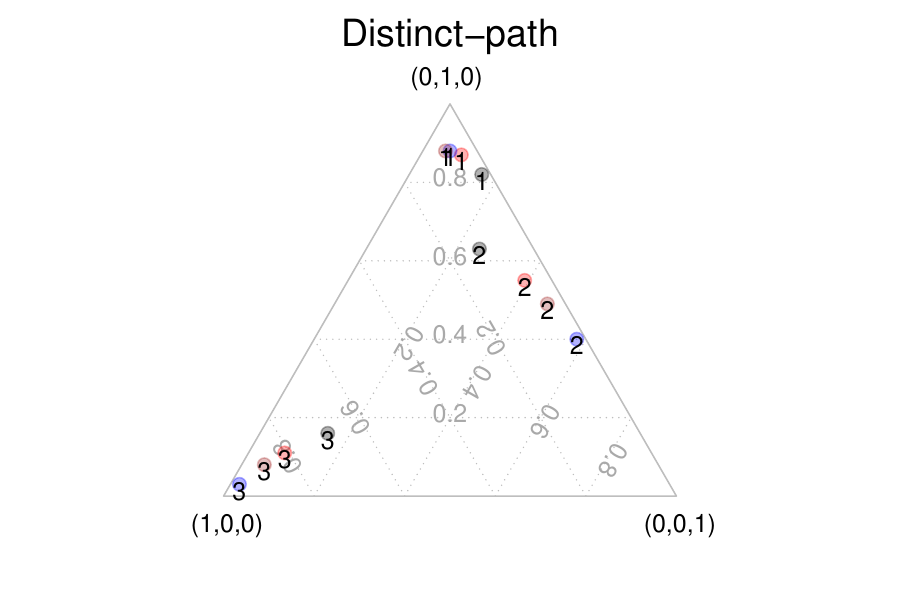} }}%
	\qquad
	\subfloat[ ]{{\includegraphics[trim = 2cm 1cm 4cm 0cm, width=6cm]{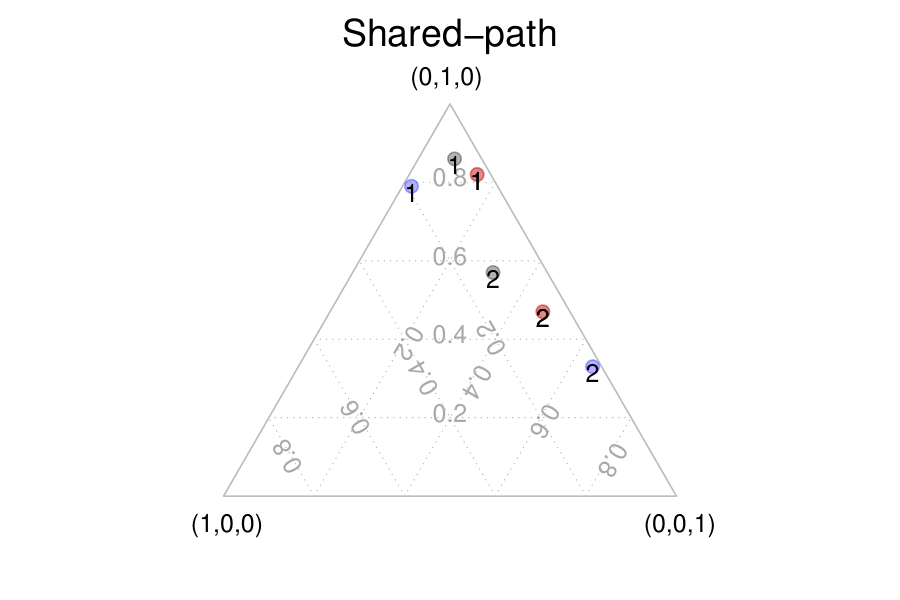} }}%
	\caption{Corresponding to Table \ref{Simu_tab_small_prob}, the Barycentric plots of different multinomial probabilities for comparing (a) distinct-path,  \textcolor{red!40}{$\boldsymbol{\bullet}$}: $\pi(A, A)$ , \textcolor{blue!40}{$\boldsymbol{\bullet}$}: $\pi(A, C)$, \textcolor{Sepia!30}{$\boldsymbol{\bullet}$}: $\pi(B, B)$, \textcolor{black!40}{$\boldsymbol{\bullet}$}: $\pi(B, E)$;
		and (b) shared-path regimes, \textcolor{red!40}{$\boldsymbol{\bullet}$}: $\pi(A, A)$ , \textcolor{blue!40}{$\boldsymbol{\bullet}$}: $\pi(A, C)$,  \textcolor{black!40}{$\boldsymbol{\bullet}$}: $\pi(A, D)$. }
	\label{Bari_plot_small_prob}%
\end{figure}

In general, the maximum likelihood estimate of a small cell probability (say less than 5\%) may end up with large bias due to less number of individuals in that cell. These biases in estimated cell probabilities make the corresponding estimated power and CP to deviate from their nominal values. Note that, the estimation problem related to a small probability/frequency is well known in the inference of categorical data \citep{Yates_1934, agresti_book}. In Table \ref{Simu_tab_small_prob}, we have considered similar scenarios to explore how our methodologies work. In all the five scenarios of Table \ref{Simu_tab_small_prob}, some of the cell probabilities are less than 5\% or close to it. In Figures \ref{Bari_plot_small_prob}(a) and \ref{Bari_plot_small_prob}(b), we have shown how the five simulation scenarios from Table \ref{Simu_tab_small_prob} are distributed in the barycentric coordinate system.

The estimated power and CP deviate from their respective values. In scenario 1 for distinct-path comparison, the estimated value of $GOR$ is almost same as the population $GOR$. Here, the estimated power and CP are not much less than their nominal values. However, in all the other four scenarios, either power or CP or both are far from their respective nominal values. The estimated power is 0.39 in scenario 3 of distinct-path comparison where the last two cell probabilities of $\pi(A, E)$ are 0.03 and 0.02, respectively. In summary, when the cell probabilities of some cells corresponding to a regime become less than 5\%, the estimated $GOR$ should be interpreted with caution.

\section{Application to SMART+ Data}\label{data_examp}

In this section, we demonstrate how the developed $GOR$ can be used to compare any two different embedded regimes using the data in SMART+ study (N=87). Note the SMART+ sample size is small, as the study was not powered for formal comparison but as a feasibility trial. We will only consider up to two stages in Figure \ref{fig.1}, as described in Section \ref{sec: smartplus}, and we shorthand the four embedded regimes as
$d(App, NC, Relaxed) = (App + Relaxed, \{App + Relaxed\}^{R_r}\{App + NC + Relaxed\}^{1-R_r}),
d(App, Relaxed) = (App + Relaxed, $ $ \{App + Relaxed\}^{R_r}\{App + Relaxed\}^{1-R_r}),
d(App, NC, Stringent) =   (App + Stringent, \{App + Stringent\}^{R_s}\{App + $ $ NC + Stringent\}^{1-R_s}),
d(App, Stringent)
= (App + Stringent, \{App + Stringent\}^{R_s}\{App + Stringent\}^{1-R_s}),
$
where $R_r$ and $R_s$ are the response indicators (1/0) following the relaxed and stringent response criteria, respectively. An athlete with a relaxed or stringent response criteria needs to have an engagement rate of at least 1 day (i.e. low response threshold) or at least 2 days (i.e. high response threshold) on the App in one week, respectively, to be considered a responder. Note that it is possible to use the $GOR$ for all three stages in Figure \ref{fig.1}, using the extended methodology for general K-stage SMART as detailed in the Supplementary Materials. Suppose we are interested in comparing clinical outcomes such as the carbohydrate periodization behavior ranking, an ordinal variable with the levels 1 (not periodizing at all), 2 (periodizing energy/kcal only) and 3 (periodizing both energy/kcal and carbohydrates); the carbohydrate periodization self-efficacy,  a 3-item measure where each statement asks the athletes to rate their confidence level on  ``{\em carbohydrate periodization}", ``{\em meal planning}" and ``{\em adherence to plan}", from 1 (not confident at all), 2 (a little confident), 3 (moderately confident), 4 (quite confident) to 5 (very confident);  and the belief about consequences, a 3-item measure about changing ``{\em health}", ``{\em performance}", and ``{\em body composition}", on a scale 1 (not at all), 2 (a confident), 3 (moderately), 4 (very much) and 5 (extremely).

In Table \ref{tab: cprank}, we present comparisons between four different pairs of distinct-path regimes and two different pairs of shared-path regimes for carbohydrate periodization behavior ranking. Here, the interpretation of $GOR_{2,1} = P(Y_{Regime^{(2)}} > Y_{Regime^{(1)}})/P(Y_{Regime^{(2)}} < Y_{Regime^{(1)}})$ is as follows: $Regime ^{(2)}$ is better than $Regime^{(1)}$ if $GOR_{2,1} > 1$, and worse if $GOR_{2,1} < 1$; a $GOR$ is significant, if the confidence intervals (CIs) do not contain 1. For the four different pairs of distinct-path regimes, we let any one of the $d(App, NC, Relaxed)$ and $d(App, Relaxed)$ be taken as $Regime^{(1)}$,  whereas any one of the $d(App, NC, Stringent)$ and $d(App, Stringent)$ can be considered as $Regime^{(2)}$; and for the two pairs of shared-path regimes, we let the regimes with NC be $Regime^{(2)}$, i.e., $d(App, Relaxed)$ is considered as $Regime^{(1)}$ when $d(App, NC, Relaxed)$ is taken as $Regime^{(2)}$. Similarly, $d(App, Stringent)$ is the $Regime^{(1)}$ when $d(App, NC,$ $ Stringent)$ is taken as $Regime^{(2)}$.


Among the four distinct-path regimes comparisons in Table \ref{tab: cprank}, regimes with stringent response criteria, $d(App, NC,$  $Stringent)$ and $d(App, Stringent)$ perform worse than regimes with relaxed response criteria, with significantly inferior performance when compared to $d(App, NC, Relaxed)$ ($GOR_{2,1}$ = 0.50 with (CI = 0.06, 0.94) and $GOR_{2,1}$ = 0.47 with CI = (0.08, 0.86), respectively). For the two different pairs of shared-path regimes comparisons, the regimes with NC, $d(App, NC, Relaxed)$ and $d(App, NC, Stringent)$, show potential superiority over their counterparts ($d(App, Relaxed)$ and $d(App, Stringent)$), respectively, although they are not statistically significant ($GOR_{2,1}$ = 1.3 with (CI = 0.24, 2.36) and $GOR_{2,1}$ = 1.08 with (CI = 0.16, 2.00), respectively). Based on these $GOR$s, one can therefore infer that using a relaxed response criteria and supporting the App with a  nutrition coach (NC) at stage 2 for non-responders are better options. It is not surprising that the regimes with NC for non-responders show better results than without NC. However, the finding that stringent response criteria is inferior to relaxed response criteria is contrary to expectation. One would normally expect that, under the stringent response criteria, athletes generally have higher chances of being assigned to receive additional support from NC early, as non-responders (e.g. an athlete with 1 day engagement would be a non-responder under the stringent response criteria, but a responder under the relaxed criteria), and hence improve their carbohydrate periodization behavior. This warrants further investigation into other aspects such as the third stage intervention in the design, the app engagement over time during the trial, the messaging frequency and interaction quality between the nutrition coaches and the athletes, which are beyond the scope of this article. As sample size were small, we highlight again the results are more indicative for future confirmatory trial than definitive.

Table \ref{tab: self-efficacy} and \ref{tab: belief} show results from the individual items in the carbohydrate periodization self-efficacy and belief about consequences measure. Note that in the SMART+ data, no athletes answered 1 (not at all) for items on beliefs about ``changing performance", and ``body composition", and hence they are effectively reduced to four categories ordinal outcomes. We give a tentative discussion of the results, considering the pilot nature of the study. Results are consistent across the self-efficacy items in Table \ref{tab: self-efficacy}, with respect to the point estimates of $GOR_{2,1}$. In Table \ref{tab: self-efficacy}, $d(App, Relaxed)$ performs the best and $d(App, NC, Relaxed)$ performs the worst among the four DTRs; between the regimes with stringent response criteria, $d(App, NC, Stringent)$ is better than $d(App, Stringent)$. However, only comparisons involving $d(App, Relaxed)$ show some significance, i.e.,  $d(App, Relaxed)$ is significantly better than $d(App, Stringent)$ across the three items; better than $d(App, NC, Stringent)$ in ``adherence to plan"; and also better than $d(App, NC, Relax)$ in ``meal planning". Note that $GOR_{2,1} > 1$ means $Regime^{(2)}$ is better than $Regime^{(1)}$, and a CI excluding 1 is significant. This indicates that giving a nutrition coach (NC) when athletes are not responding (i.e. have low engagement), does not help improve self-efficacy, especially in ``adherence to plan" and ``meal planning".  Results on beliefs about consequences (Table \ref{tab: belief}) are less consistent among the three items. For the item on ``health", the $GOR$s show it is better to have no NC (i.e. $d(App, Relaxed)$, and likewise for $d(App, Stringent)$, is better than $d(App, NC, Relaxed)$ and $d(App, NC, stringent)$) and use a stringent response criteria (i.e.  $d(App, Stringent)$ is better than $d(App, Relaxed)$; $d(App, NC, Stringent)$ is better than $d(App, NC, Relaxed)$). However, none of these comparisons are significant. The results are identical between items ``performance" and ``body composition", where the interpretation is the same as for self-efficacy, except none of the comparisons are significant.


\section{Discussion}\label{discussion}

\begin{figure}[ht]
	\centering
	\includegraphics[trim =6cm 0cm 8cm 0cm, height=8cm, width=10cm]{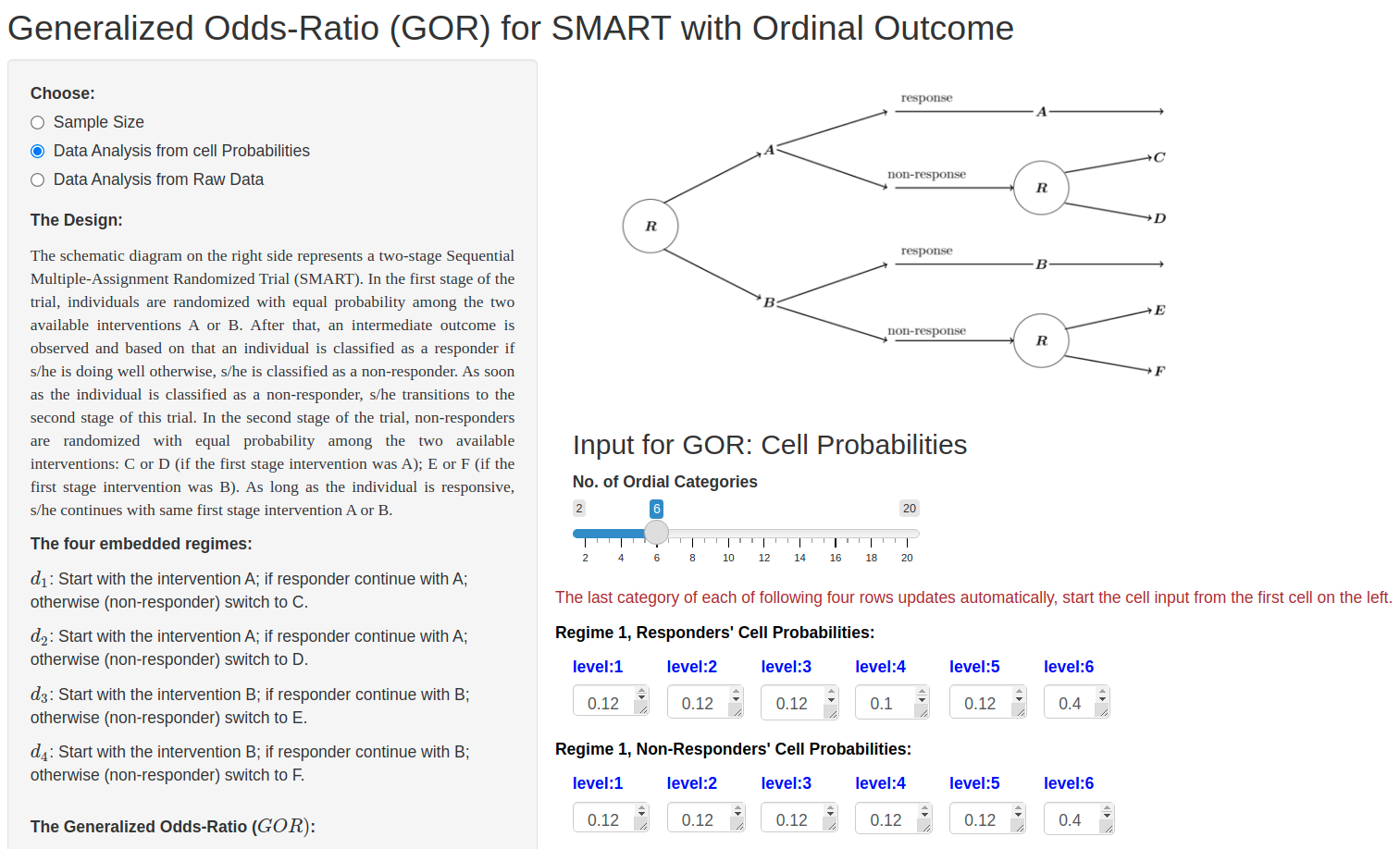}  
	\caption{A screenshot of the developed interactive Shiny web app using R. It provides, i) sample size estimation, ii) data analysis using only cell probabilities from a completed study, and iii) data analysis from raw data in Excel format.}
	\label{fig.shinyapp}
\end{figure}

The SMART is a very flexible design that is not limited to the structure in Figure \ref{fig.1}. In Figure \ref{fig.1}, responders are not re-randomized; \citet{Nahum-Shani2012a} refers to this type of SMARTs as a ``restricted" SMART. There are also many other variations, such as the ``unrestricted" SMART where both responders and non-responders are re-randomized potentially to different interventions \citep{Nahum2019_education}. As personal data collection becomes increasingly easy from mobile devices and sensors, the notion of developing DTRs and using SMARTs is progressively gaining traction across many areas of clinical sciences. In this context, we view the introduction of the generalized odds ratio ($GOR$) as timely, as the demand for more specialised methodologies to deal with various types of SMART-related outcome measures (e.g. binary, ordinal, continuous, composite and survival) grows in conjunction with the increasing number of SMARTs in the field. The $GOR$ is a useful measure of association that is suitable for binary, ordinal and continuous outcomes from SMART. The $GOR$ is also easy to interpret as it is an extension of the simple odds ratio that is widely used in clinical and behavioral sciences. A freely available Shiny web app (\textcolor{blue}{\url{https://www.iitg.ac.in/pgapps/dGOR/}}) using R is provided to make the proposed method accessible to other researchers and practitioners. Figure \ref{fig.shinyapp} shows a screenshot of the developed Shiny web app. The web app can perform sample size estimation, data analysis using only cell probabilities from a completed study, and data analysis from raw data in Excel format.



The $GOR$ specifically fills the literature gap in the estimation and comparison of embedded DTRs in SMARTs for ordinal outcomes. We highlight the value of providing this methodology by using the SMART+ study as an example. In the SMART+ study, the primary end-point was defined as the success in attaining carbohydrate periodization behavior, by collapsing the ordinal scale, such that 1 (not periodizing at all) and 2 (periodizing energy/kcal only) are considered failure (i.e. success = 0) and 3 (periodizing both energy/kcal and carbohydrates) as success. Note that dichotomising continuous or ordinal measures is a common yet problematic practice in clinical research \citep{altman2006}; conventional studies, however, can work around this using well established methods, such as the standard $GOR$ \citep{Agresti1980}. If we combine the outcome cell probabilities in Table \ref{tab: cprank} into two categories, success = 1/0 as defined earlier, the regime $d(App, Stringent)$ will be identical to $d(App, NC, Stringent)$, because all the non-responders in both regimes are now categorized as failure (i.e., both have outcome cell probabilities (0.769, 0.231) and (1,0) for responders amd non-responders, respectively). The comparison between the regimes pair becomes futile, as valuable information on the intermediate level is now lost. One limitation of the $GOR$ is that, the estimates may be biased when at least some of the cell probabilities corresponding to any of the two regimes under comparison are small (say, $< 0.05$ or so), as discussed in Section \ref{simu_small_prob}. However, this phenomenon is not unique to $GOR$, but happens in many classical estimation problems involving categorical data \citep{agresti_book}. 


\section*{Supplementary Materials}
The supplementary materials include some detailed calculations related to $GOR$ in Section 1; proofs of Propositions 1 and 2 in Section 2; likelihood for two-stage SMART in Section 3; $GOR$ for $K$-stage SMART in Section 4; a simulation study in distinct-path regimes with different response rates in Section 5; discussion on ``unrestricted" SMART with simulation in Section 6; asymptotic distribution of $GOR$ for distinct-path comparison for two-stage and $K$-stage SMART in Section 7 and 8, for  shared-path comparison in Section 9;  and data generation algorithm for SMART with an ordinal outcome in Section 10.

\section*{Acknowledgments}
Dr Bibhas Chakraborty would like to acknowledge support by Khoo Bridge Funding Award from the Duke-NUS Medical School (grant number: Duke-NUS-KBrFA/2021/0040). Xiaoxi Yan would like to acknowledge funding support by Duke-NUS Medical School as part of her PhD.

\section*{Conflict of Interest}
Xiaoxi Yan is one of the founders of Applied Behaviour Systems Ltd. Hexis Performance is developed by Applied Behaviour Systems Ltd.

\bibliographystyle{chicago}
\bibliography{refs}

\setlength\tabcolsep{3pt} 

\begin{sidewaystable}[ph!]
	\centering
	
	\caption{Comparison of Distinct-path regimes with an ordinal outcome. Here, $\gamma_{A} = 0.3$ and $\gamma_{B}=0.4$,  nominal power is 0.80 and type-I error rate is 0.05. The standardized effect size is in log-scale. Estimated powers and coverage probabilities are based on 5,000 simulations.} 
	\begin{tabular}{ccccccccccc}
		Serial number & $\pi(A,A)$ & $\pi(B,B)$ & Std. ES &
		$\eta_{d^{(3)},d^{(1)}}^{\text{DP}}$ & $N$ & $\hat\eta_{d^{(3)},d^{(1)}}^{\text{DP}}$ & $SSE$ & $ASE$ & $\widehat{power}$ & $\widehat{CP}$ \\ 
		& $\pi(A,C)$ & $\pi(B,E)$ & (log scale) & & & & & & & \\ 
		&&& &&&& &&& \\
		1 & (0.23, 0.51, 0.26) &  (0.31, 0.50, 0.19) & & & & & & & & \\ 
		& (0.50, 0.41, 0.09) & (0.14, 0.47, 0.39) & 0.219 & 2.55 & 164 & 2.72 & 1.06 & 1.14 & 0.78 & 0.94 \\
		&&& &&&& &&& \\
		2 & (0.41, 0.23, 0.36) &  (0.50, 0.22, 0.28) & & & & & & & & \\ 
		&  (0.58, 0.20, 0.22) &  (0.27, 0.22, 0.51) & 0.147 & 1.86 & 366 & 1.89 & 0.43 & 0.43 & 0.78 & 0.95 \\ 
		&&& &&&& &&& \\
		3 &  (0.21, 0.40, 0.39) &  (0.28, 0.41, 0.31) & & & & & & & & \\ 
		&  (0.30, 0.41, 0.29) &  (0.12, 0.34, 0.54) & 0.117 & 1.64 & 571 & 1.66 & 0.29 & 0.30 & 0.79 & 0.95 \\
		&&& &&&& &&& \\ 
		4 &   (0.13, 0.22, 0.65) &  (0.10, 0.19, 0.71) & & & & & & & & \\ 
		&   (0.09, 0.18, 0.73) &  (0.20, 0.26, 0.54) & -0.085 & 0.66 & 1096 & 0.67 & 0.10 & 0.10 & 0.78 & 0.95 \\
		&&& &&&& &&& \\ 
		5 &  (0.12, 0.24, 0.64) &  (0.09, 0.21, 0.70) & & & & & & & & \\ 
		&  (0.07, 0.18, 0.75) &  (0.18, 0.28, 0.54) & -0.099 & 0.61 & 797 & 0.61 & 0.12 & 0.11 & 0.82 & 0.95 \\ 
		&&& &&&& &&& \\ 
		6 &  (0.28, 0.52, 0.20) &  (0.34, 0.50, 0.16) & & & & & & & & \\ 
		&  (0.08, 0.43, 0.49) &  (0.20, 0.52, 0.28) & -0.161 & 0.50 & 305 & 0.53 & 0.26 & 0.16 & 0.81 & 0.94 \\
		
		&&& &&&& &&& \\   
	\end{tabular}
	\label{Simu_tab_DP}
	
\end{sidewaystable}

\setlength\tabcolsep{3pt} 

\begin{sidewaystable}[ph!]
	\centering
	\caption{Comparison of Shared-path regimes with an ordinal outcome. Here, $\gamma_{A} = 0.3$ and $\gamma_{B}=0.4$,  nominal power is 0.80 and type-I error rate is 0.05. The standardized effect size is in log-scale. Estimated powers and coverage probabilities are based on 5,000 simulations.} 
	\begin{tabular}{ccccccccccc}
		Serial number & \multicolumn{2}{c}{$\pi(A,A)$} & Std. ES & $\eta_{d^{(2)},d^{(1)}}^{\text{SP}}$ & $N$ & $\hat\eta_{d^{(2)},d^{(1)}}^{\text{SP}}$ & $SSE$ & $ASE$ & $\widehat{power}$ & $\widehat{CP}$\\ 
		& $\pi(A,C)$ & $\pi(A,D)$ & (log scale) & &&& &&& \\ 
		&& & &&& &&& & \\
		1 & \multicolumn{2}{c}{(0.24, 0.52, 0.24)} && &&& &&& \\ 
		& (0.63, 0.33, 0.04) & (0.38, 0.49, 0.13) & 0.161 & 1.88 & 304 & 1.94 & 0.44 & 0.49 & 0.77 & 0.97 \\ 
		&& & &&& &&& & \\
		2 & \multicolumn{2}{c}{(0.03, 0.66, 0.31)} && &&& &&& \\ 
		& (0.19, 0.74, 0.07) & (0.07, 0.75, 0.18) & 0.148 & 1.96 & 357 & 1.97 & 0.54 & 0.51 & 0.77 & 0.94 \\
		&& & &&& &&& & \\
		3 & \multicolumn{2}{c}{(0.43, 0.17, 0.40)} && &&& &&& \\ 
		& (0.73, 0.12, 0.15) & (0.56, 0.16, 0.28) & 0.109 & 1.56 & 659 & 1.60 & 0.26 & 0.27 & 0.81 & 0.96 \\ 
		&& & &&& &&& & \\
		4 & \multicolumn{2}{c}{(0.36, 0.36, 0.28)} && &&& &&& \\ 
		& (0.40, 0.35, 0.25) & (0.52, 0.32, 0.16) & -0.080 & 0.73 & 1218 & 0.73 & 0.08 & 0.09 & 0.80 & 0.95 \\ 
		&& & &&& &&& & \\
		5 &  \multicolumn{2}{c}{(0.17, 0.10, 0.73)} && &&& &&& \\ 
		
		& (0.15, 0.10, 0.75) & (0.29, 0.13, 0.58) & -0.113 & 0.58 & 618 & 0.58 & 0.12 & 0.12 & 0.82 & 0.94 \\
		&& & &&& &&& & \\
		6 & \multicolumn{2}{c}{(0.24, 0.35, 0.41)} && &&& &&& \\ 
		& (0.16, 0.32, 0.52) & (0.38, 0.35, 0.27) & -0.181 & 0.50 & 241 & 0.50 & 0.13 & 0.14 & 0.80 & 0.96 \\ 
	\end{tabular}
	\label{Simu_tab_SP}
\end{sidewaystable} 


\setlength\tabcolsep{3pt} 

\begin{sidewaystable}[ph!]
	\centering
	\begin{small}
		\caption{Some examples with small cell probabilities. Here, $\gamma_{A} = 0.3$ and $\gamma_{B}=0.4$,  nominal power is 0.80 and type-I error rate is 0.05. The standardized effect size is in log-scale. Estimated powers and coverage probabilities are based on 5,000 simulations. } 
		\begin{tabular}{ccccccccccc}
			$SN$ & $\pi(A,A)$ & $\pi(B,B)$ & Std. ES & $\eta_{d^{(3)},d^{(1)}}^{\text{DP}}$ & $N$ & $\hat\eta_{d^{(3)},d^{(1)}}^{\text{DP}}$ & $SSE$ & $ASE$ & $\widehat{power}$ & $\widehat{CP}$ \\ 
			& $\pi(A,C)$ & $\pi(B,E)$ & (log scale) & & & & & & & \\ 
			&&& &&&& &&& \\
			& Comparison of distinct-path: && &&&& &&& \\
			&&& &&&& &&& \\
			1 & (0.04, 0.87, 0.09) &  (0.07, 0.88, 0.05) & & & & & & & & \\ 
			& (0.06, 0.88, 0.06) & (0.02, 0.82, 0.16) & 0.072 & 1.68 & 1506 & 1.67 & 0.39 & 0.32 & 0.77 & 0.92  \\
			&&& &&&& &&& \\
			2 & (0.06, 0.55, 0.39) &  (0.04, 0.49, 0.47) & & & & & & & & \\ 
			& (0.02, 0.40, 0.58)  & (0.12, 0.63, 0.26) & -0.156 & 0.48 & 322 & 1.73 & 2.63 & 3.29 & 0.79 & 0.56 \\
			&&& &&&& &&& \\
			3 & (0.81, 0.11, 0.08) & (0.87, 0.08, 0.05) & & & & & & & & \\ 
			& (0.95, 0.03, 0.02) & (0.69, 0.16, 0.15) & 0.170 & 3.23 & 272 & 3.63 & 2.06 & 10.7 & 0.39 & 0.97 \\ 
			&&& &&&& &&& \\
			& Comparison of shared-path: && &&&& &&& \\
			&&& &&&& &&& \\
			& \multicolumn{2}{c}{$\pi(A,A)$} &  & $\eta_{d^{(2)},d^{(1)}}^{\text{SP}}$ &  & $\hat\eta_{d^{(2)},d^{(1)}}^{\text{SP}}$ & & && \\ 
			& $\pi(A,E)$ & $\pi(A,F)$ &  & &&& &&& \\ 
			&&& &&&& &&& \\
			1 & \multicolumn{2}{c}{(0.03, 0.82, 0.15)} & & & & & & & & \\ 
			& (0.19, 0.79, 0.02) & (0.06, 0.86, 0.08) & 0.137 & 2.23 & 420 & 2.16 & 0.74 & 0.71 & 0.74 & 0.87 \\
			&&& &&&& &&& \\ 
			2 &  \multicolumn{2}{c}{(0.06, 0.47, 0.47)}& & & & & & & & \\ 
			&  (0.02, 0.33, 0.65) & (0.12, 0.57, 0.31) & -0.219 & 0.38 & 164 & 1.87 & 1.43 & 1.16 & 0.74 & 0.43 \\
			&&& &&&& &&& \\
		\end{tabular}
		\label{Simu_tab_small_prob}
	\end{small}
\end{sidewaystable}

\begin{sidewaystable}[ph!]
	\centering
	\caption{Comparison of regimes with carbohydrate periodization behavior ranking as the ordinal outcome in SMART+ data. The response rates are $\gamma_R = 0.64$ and $\gamma_S = 0.52$, following the first-stage intervention with ``Relaxed" and ``Stringent" response criteria, respectively. Note that, $GOR_{2,1} = P(Y_{Regime^{(2)}} > Y_{Regime^{(1)}})/P(Y_{Regime^{(2)}} < Y_{Regime^{(1)}})$.}
	\begin{tabular}{llllll}
		&&& && \\
		\multicolumn{1}{c}{$Regime^{(1)}$}& Outcome cell  & \multicolumn{1}{c}{$Regime^{(2)}$} & Outcome cell & $GOR_{2,1}$ & CI.$GOR_{2,1}$ \\
		& probabilities: && probabilities: && \\
		& (Responder) &  & (Responder) & &  \\
		& (Non-Responder) &  & (Non-Responder) & &  \\
		&&& && \\
		\multicolumn{6}{c}{Comparison of distinct-path regimes}  \\
		&&& && \\
		$d(App, NC, Relaxed)$ & ( 0.360, 0.360, 0.280 ) & $d(App, NC, Stringent)$ & ( 0.500, 0.269, 0.231 ) & 0.50 &  ( 0.06, 0.94 ) \\  
		& ( 0.500, 0.125, 0.375 ) &   & ( 0.615, 0.385, 0 ) &  &  \\ 
		&&& && \\
		$d(App, NC, Relaxed)$ & ( 0.360, 0.360, 0.280 ) & $d(App, Stringent)$ & ( 0.500, 0.269, 0.231 ) & 0.47 &  ( 0.08, 0.86 ) \\   
		& ( 0.500, 0.125, 0.375 ) &   & ( 0.667, 0.333, 0 ) &   &  \\ 
		&&& && \\
		$d(App, Relaxed)$ & ( 0.360, 0.360, 0.280 ) & $d(App, NC, Stringent)$ & ( 0.500, 0.269, 0.231 ) & 0.66 &  ( 0.14, 1.18 ) \\   
		& ( 0.667, 0.167, 0.166 ) &   & ( 0.615, 0.385, 0 ) &  &    \\ 
		&&& && \\
		$d(App, Relaxed)$ & ( 0.360, 0.360, 0.280 ) & $d(App, Stringent)$ & ( 0.500, 0.269, 0.231 ) & 0.61 &  ( 0.09, 1.13 ) \\
		& ( 0.667, 0.167, 0.166 ) &   & ( 0.667, 0.333, 0 ) &  &    \\
		&&& && \\
		
		\multicolumn{6}{c}{Comparison of shared-path regimes}  \\
		&&& && \\
		$d(App, Relaxed)$ & ( 0.360, 0.360, 0.280 ) & $d(App, NC, Relaxed)$ & ( 0.360, 0.360, 0.280 ) & 1.30 &  ( 0.24, 2.36 ) \\   
		& ( 0.667, 0.167, 0.166 ) &   & ( 0.500, 0.125, 0.375 ) &   &  \\ 
		&&& && \\
		$d(App, Stringent)$ & ( 0.500, 0.269, 0.231 ) & $d(App, NC, Stringent)$ & ( 0.500, 0.269, 0.231 ) & 1.08 &  ( 0.16, 2.00 ) \\   
		& ( 0.667, 0.333, 0 )  &   & ( 0.615, 0.385, 0 ) &  &  \\
		&&& && \\
	\end{tabular}
	\label{tab: cprank}
\end{sidewaystable} 

\begin{table}[ph!] 
	\centering
	\caption{Comparison of regimes with self-efficacy items (a) carbohydrate periodization, (b) meal planning and (c) adherence to plan, from SMART+ data. The response rates are $\gamma_R = 0.64$ and $\gamma_S = 0.52$, following the first-stage intervention with ``Relaxed" and ``Stringent" response criteria, respectively. Note that, $GOR_{2,1} = P(Y_{Regime^{(2)}} > Y_{Regime^{(1)}})/P(Y_{Regime^{(2)}} < Y_{Regime^{(1)}})$. }
	\resizebox{\textwidth}{!}{
		\begin{tabular}{llllll}
			&&& && \\
			\hline
			\multicolumn{1}{c}{$Regime^{(1)}$}& Outcome cell  & \multicolumn{1}{c}{$Regime^{(2)}$} & Outcome cell & $GOR_{2,1}$ & CI.$GOR_{2,1}$ \\
			& probabilities: && probabilities: && \\
			& (Responder) &  & (Responder) & &  \\
			& (Non-Responder) &  & (Non-Responder) & &  \\
			\hline
			\multicolumn{6}{c}{(a) carbohydrate periodization}  \\
			\hline
			\multicolumn{6}{c}{Comparison of distinct-path regimes}  \\
			&&& && \\
			$d(App, Relaxed)$ & ( 0, 0.24, 0.36, 0.4, 0 ) & $d(App, Stringent)$ & ( 0.08, 0.12, 0.39, 0.23, 0.19 ) & 0.56 & ( 0.17, 0.95 ) \\
			& ( 0, 0, 0.33, 0.33, 0.33 ) & & ( 0.22, 0.11, 0.44, 0.22, 0 ) & & \\
			$d(App, Relaxed)$ & ( 0, 0.24, 0.36, 0.4, 0 ) & $d(App, NC, Stringent)$ & ( 0.08, 0.12, 0.39, 0.23, 0.19 ) & 0.65 & ( 0.21, 1.09 ) \\
			& ( 0, 0, 0.33, 0.33, 0.33 ) & & ( 0, 0.15, 0.69, 0.08, 0.08 ) & & \\
			$d(App, NC, Relaxed)$ & ( 0, 0.24, 0.36, 0.4, 0 ) & $d(App,  Stringent)$ & ( 0.08, 0.12, 0.39, 0.23, 0.19 ) & 1.01 & ( 0.28, 1.74 ) \\
			& ( 0.13, 0.38, 0.13, 0.38, 0 ) & & ( 0.22, 0.11, 0.44, 0.22, 0 ) & & \\
			$d(App, NC, Relaxed)$ & ( 0, 0.24, 0.36, 0.4, 0 ) & $d(App,  NC, Stringent)$ & ( 0.08, 0.12, 0.39, 0.23, 0.19 ) & 1.24 & ( 0.34, 2.14 ) \\
			& ( 0.13, 0.38, 0.13, 0.38, 0 ) & & ( 0, 0.15, 0.69, 0.08, 0.08 ) & & \\
			&&& && \\
			
			\multicolumn{6}{c}{Comparison of shared-path regimes}  \\
			&&& && \\
			$d(App, Relaxed)$ & ( 0, 0.24, 0.36, 0.4, 0 ) & $d(App,  NC, Relaxed)$ & ( 0, 0.24, 0.36, 0.4, 0 ) & 0.61 & ( 0.13, 1.09 ) \\
			& ( 0, 0, 0.33, 0.33, 0.33 ) & & ( 0.13, 0.38, 0.13, 0.38, 0 ) & & \\
			$d(App, Stringent)$ & ( 0.08, 0.12, 0.39, 0.23, 0.19 ) & $d(App,  NC, Stringent)$ & ( 0.08, 0.12, 0.39, 0.23, 0.19 ) & 1.2 & ( 0.32, 2.08 ) \\
			& ( 0.22, 0.11, 0.44, 0.22, 0 ) & & ( 0, 0.15, 0.69, 0.08, 0.08 ) & & \\
			\hline
			
			\multicolumn{6}{c}{(b) meal planning}  \\
			\hline
			\multicolumn{6}{c}{Comparison of distinct-path regimes}  \\
			&&& && \\
			$d(App, Relaxed)$ & ( 0, 0.12, 0.48, 0.4, 0 ) & $d(App, Stringent)$ & ( 0.04, 0.23, 0.31, 0.27, 0.15 ) & 0.51 & ( 0.12, 0.9 ) \\
			& ( 0, 0.17, 0, 0.5, 0.33 ) & & ( 0.11, 0.22, 0.33, 0.33, 0 ) & & \\
			$d(App, Relaxed)$ & ( 0, 0.12, 0.48, 0.4, 0 ) & $d(App, NC, Stringent)$ & ( 0.04, 0.23, 0.31, 0.27, 0.15 ) & 0.7 & ( 0.18, 1.22 ) \\
			& ( 0, 0.17, 0, 0.5, 0.33 ) & & ( 0, 0.15, 0.39, 0.39, 0.08 ) & & \\
			$d(App, NC, Relaxed)$ & ( 0, 0.12, 0.48, 0.4, 0 ) & $d(App,  Stringent)$ & ( 0.04, 0.23, 0.31, 0.27, 0.15 ) & 1.15 & ( 0.27, 2.03 ) \\
			& ( 0.13, 0.25, 0.63, 0, 0 ) & & ( 0.11, 0.22, 0.33, 0.33, 0 ) & & \\
			$d(App, NC, Relaxed)$ & ( 0, 0.12, 0.48, 0.4, 0 ) & $d(App,  NC, Stringent)$ & ( 0.04, 0.23, 0.31, 0.27, 0.15 ) & 1.69 & ( 0.42, 2.96 ) \\
			& ( 0.13, 0.25, 0.63, 0, 0 ) & & ( 0, 0.15, 0.39, 0.39, 0.08 ) & & \\
			&&& && \\
			
			\multicolumn{6}{c}{Comparison of shared-path regimes}  \\
			&&& && \\
			$d(App, Relaxed)$ & ( 0, 0.12, 0.48, 0.4, 0 ) & $d(App,  NC, Relaxed)$ & ( 0, 0.12, 0.48, 0.4, 0 ) & 0.36 & ( 0.08, 0.64 ) \\
			& ( 0, 0.17, 0, 0.5, 0.33 ) & & ( 0.13, 0.25, 0.63, 0, 0 ) & & \\
			$d(App, Stringent)$ & ( 0.04, 0.23, 0.31, 0.27, 0.15 ) & $d(App,  NC, Stringent)$ & ( 0.04, 0.23, 0.31, 0.27, 0.15 ) & 1.37 & ( 0.35, 2.39 ) \\
			& ( 0.11, 0.22, 0.33, 0.33, 0 ) & & ( 0, 0.15, 0.39, 0.39, 0.08 ) & & \\
			\hline
			\multicolumn{6}{c}{(c) adherence to plan}  \\
			\hline
			\multicolumn{6}{c}{Comparison of distinct-path regimes}  \\
			&&& && \\
			$d(App, Relaxed)$ & ( 0.04, 0.2, 0.4, 0.28, 0.08 ) & $d(App, Stringent)$ & ( 0.04, 0.23, 0.23, 0.31, 0.19 ) & 0.59 & ( 0.2, 0.98 ) \\
			& ( 0, 0.17, 0, 0.5, 0.33 ) & & ( 0.22, 0.11, 0.44, 0.22, 0 ) & & \\
			$d(App, Relaxed)$ & ( 0.04, 0.2, 0.4, 0.28, 0.08 ) & $d(App, NC, Stringent)$ & ( 0.04, 0.23, 0.23, 0.31, 0.19 ) & 0.59 & ( 0.2, 0.98 ) \\
			& ( 0, 0.17, 0, 0.5, 0.33 ) & & ( 0.08, 0.23, 0.54, 0.15, 0 ) & & \\
			$d(App, NC, Relaxed)$ & ( 0.04, 0.2, 0.4, 0.28, 0.08 ) & $d(App,  Stringent)$ & ( 0.04, 0.23, 0.23, 0.31, 0.19 ) & 1.04 & ( 0.28, 1.8 ) \\
			& ( 0.13, 0.38, 0.25, 0, 0.25 ) & & ( 0.22, 0.11, 0.44, 0.22, 0 ) & & \\
			$d(App, NC, Relaxed)$ & ( 0.04, 0.2, 0.4, 0.28, 0.08 ) & $d(App,  NC, Stringent)$ & ( 0.04, 0.23, 0.23, 0.31, 0.19 ) & 1.07 & ( 0.29, 1.85 ) \\
			& ( 0.13, 0.38, 0.25, 0, 0.25 ) & & ( 0.08, 0.23, 0.54, 0.15, 0 ) & & \\
			&&& && \\
			
			\multicolumn{6}{c}{Comparison of shared-path regimes}  \\
			&&& && \\
			$d(App, Relaxed)$ & ( 0.04, 0.2, 0.4, 0.28, 0.08 ) & $d(App,  NC, Relaxed)$ & ( 0.04, 0.2, 0.4, 0.28, 0.08 ) & 0.58 & ( 0.14, 1.02 ) \\
			& ( 0, 0.17, 0, 0.5, 0.33 ) & & ( 0.13, 0.38, 0.25, 0, 0.25 ) & & \\
			$d(App, Stringent)$ & ( 0.04, 0.23, 0.23, 0.31, 0.19 ) & $d(App,  NC, Stringent)$ & ( 0.04, 0.23, 0.23, 0.31, 0.19 ) & 1.02 & ( 0.31, 1.73 ) \\
			& ( 0.22, 0.11, 0.44, 0.22, 0 ) & & ( 0.08, 0.23, 0.54, 0.15, 0 ) & & \\
			\hline	
	\end{tabular}}
	\label{tab: self-efficacy}
\end{table}

\begin{table}
	\centering
	\caption{Comparison of regimes with belief about consequences items (a) health, (b) performance and (c) body composition, from SMART+ data. The response rates are $\gamma_R = 0.64$ and $\gamma_S = 0.52$, following the first-stage intervention with ``Relaxed" and ``Stringent" response criteria, respectively. Note that, $GOR_{2,1} = P(Y_{Regime^{(2)}} > Y_{Regime^{(1)}})/P(Y_{Regime^{(2)}} < Y_{Regime^{(1)}})$. }
	\resizebox{\textwidth}{!}{
		\begin{tabular}{llllll}
			&&& && \\
			\hline
			\multicolumn{1}{c}{$Regime^{(1)}$}& Outcome cell  & \multicolumn{1}{c}{$Regime^{(2)}$} & Outcome cell & $GOR_{2,1}$ & CI.$GOR_{2,1}$ \\
			& probabilities: && probabilities: && \\
			& (Responder) &  & (Responder) & &  \\
			& (Non-Responder) &  & (Non-Responder) & &  \\
			\hline
			\multicolumn{6}{c}{(a) health}  \\
			\hline
			\multicolumn{6}{c}{Comparison of distinct-path regimes}  \\
			&&& && \\
			$d(App, Relaxed)$ & ( 0.04, 0.16, 0.48, 0.2, 0.12 ) & $d(App, Stringent)$ & ( 0, 0.08, 0.27, 0.42, 0.23 ) & 1.2 & ( 0.32, 2.08 ) \\
			& ( 0, 0.17, 0.17, 0.17, 0.5 ) & & ( 0.11, 0, 0.44, 0.33, 0.11 ) & & \\
			$d(App, Relaxed)$ & ( 0.04, 0.16, 0.48, 0.2, 0.12 ) & $d(App, NC, Stringent)$ & ( 0, 0.08, 0.27, 0.42, 0.23 ) & 0.77 & ( 0.25, 1.29 ) \\
			& ( 0, 0.17, 0.17, 0.17, 0.5 ) & & ( 0.08, 0.31, 0.46, 0.15, 0 ) & & \\
			$d(App, NC, Relaxed)$ & ( 0.04, 0.16, 0.48, 0.2, 0.12 ) & $d(App,  Stringent)$ & ( 0, 0.08, 0.27, 0.42, 0.23 ) & 1.93 & ( 0.4, 3.46 ) \\
			& ( 0, 0.25, 0.38, 0.25, 0.13 ) & & ( 0.11, 0, 0.44, 0.33, 0.11 ) & & \\
			$d(App, NC, Relaxed)$ & ( 0.04, 0.16, 0.48, 0.2, 0.12 ) & $d(App,  NC, Stringent)$ & ( 0, 0.08, 0.27, 0.42, 0.23 ) & 1.12 & ( 0.34, 1.9 ) \\
			& ( 0, 0.25, 0.38, 0.25, 0.13 ) & & ( 0.08, 0.31, 0.46, 0.15, 0 ) & & \\
			&&& && \\
			
			\multicolumn{6}{c}{Comparison of shared-path regimes}  \\
			&&& && \\
			$d(App, Relaxed)$ & ( 0.04, 0.16, 0.48, 0.2, 0.12 ) & $d(App,  NC, Relaxed)$ & ( 0.04, 0.16, 0.48, 0.2, 0.12 ) & 0.69 & ( 0.17, 1.21 ) \\
			& ( 0, 0.17, 0.17, 0.17, 0.5 ) & & ( 0, 0.25, 0.38, 0.25, 0.13 ) & & \\
			$d(App, Stringent)$ & ( 0, 0.08, 0.27, 0.42, 0.23 ) & $d(App,  NC, Stringent)$ & ( 0, 0.08, 0.27, 0.42, 0.23 ) & 0.6 & ( 0.16, 1.04 ) \\
			& ( 0.11, 0, 0.44, 0.33, 0.11 ) & & ( 0.08, 0.31, 0.46, 0.15, 0 ) & & \\
			\hline
			
			\multicolumn{6}{c}{(b) performance}  \\
			\hline
			\multicolumn{6}{c}{Comparison of distinct-path regimes}  \\
			&&& && \\
			$d(App, Relaxed)$ & ( 0.12, 0.28, 0.4, 0.2 ) & $d(App, Stringent)$ & ( 0.08, 0.15, 0.39, 0.39 ) & 0.65 & ( 0.21, 1.09 ) \\
			& ( 0, 0.17, 0.17, 0.67 ) & & ( 0.11, 0.56, 0.22, 0.11 ) & & \\
			$d(App, Relaxed)$ & ( 0.12, 0.28, 0.4, 0.2 ) & $d(App, NC, Stringent)$ & ( 0.08, 0.15, 0.39, 0.39 ) & 0.7 & ( 0.22, 1.18 ) \\
			& ( 0, 0.17, 0.17, 0.67 ) & & ( 0.23, 0.23, 0.46, 0.08 ) & & \\
			$d(App, NC, Relaxed)$ & ( 0.12, 0.28, 0.4, 0.2 ) & $d(App,  Stringent)$ & ( 0.08, 0.15, 0.39, 0.39 ) & 1.06 & ( 0.3, 1.82 ) \\
			& ( 0.25, 0.13, 0.38, 0.25 ) & & ( 0.11, 0.56, 0.22, 0.11 ) & & \\
			$d(App, NC, Relaxed)$ & ( 0.12, 0.28, 0.4, 0.2 ) & $d(App,  NC, Stringent)$ & ( 0.08, 0.15, 0.39, 0.39 ) & 1.14 & ( 0.29, 1.99 ) \\
			& ( 0.25, 0.13, 0.38, 0.25 ) & & ( 0.23, 0.23, 0.46, 0.08 ) & & \\
			&&& && \\
			
			\multicolumn{6}{c}{Comparison of shared-path regimes}  \\
			&&& && \\
			$d(App, Relaxed)$ & ( 0.12, 0.28, 0.4, 0.2 ) & $d(App,  NC, Relaxed)$ & ( 0.12, 0.28, 0.4, 0.2 ) & 0.62 & ( 0.18, 1.06 ) \\
			& ( 0, 0.17, 0.17, 0.67 ) & & ( 0.25, 0.13, 0.38, 0.25 ) & & \\
			$d(App, Stringent)$ & ( 0.08, 0.15, 0.39, 0.39 ) & $d(App,  NC, Stringent)$ & ( 0.08, 0.15, 0.39, 0.39 ) & 1.08 & ( 0.32, 1.84 ) \\
			& ( 0.11, 0.56, 0.22, 0.11 ) & & ( 0.23, 0.23, 0.46, 0.08 ) & & \\
			\hline
			\multicolumn{6}{c}{(c) body composition}  \\
			\hline
			\multicolumn{6}{c}{Comparison of distinct-path regimes}  \\
			&&& && \\
			$d(App, Relaxed)$ & ( 0.12, 0.28, 0.4, 0.2 ) & $d(App, Stringent)$ & ( 0.08, 0.15, 0.39, 0.39 ) & 0.65 & ( 0.21, 1.09 ) \\
			& ( 0, 0.17, 0.17, 0.67 ) & & ( 0.11, 0.56, 0.22, 0.11 ) & & \\
			$d(App, Relaxed)$ & ( 0.12, 0.28, 0.4, 0.2 ) & $d(App, NC, Stringent)$ & ( 0.08, 0.15, 0.39, 0.39 ) & 0.7 & ( 0.22, 1.18 ) \\
			& ( 0, 0.17, 0.17, 0.67 ) & & ( 0.23, 0.23, 0.46, 0.08 ) & & \\
			$d(App, NC, Relaxed)$ & ( 0.12, 0.28, 0.4, 0.2 ) & $d(App,  Stringent)$ & ( 0.08, 0.15, 0.39, 0.39 ) & 1.06 & ( 0.3, 1.82 ) \\
			& ( 0.25, 0.13, 0.38, 0.25 ) & & ( 0.11, 0.56, 0.22, 0.11 ) & & \\
			$d(App, NC, Relaxed)$ & ( 0.12, 0.28, 0.4, 0.2 ) & $d(App,  NC, Stringent)$ & ( 0.08, 0.15, 0.39, 0.39 ) & 1.14 & ( 0.29, 1.99 ) \\
			& ( 0.25, 0.13, 0.38, 0.25 ) & & ( 0.23, 0.23, 0.46, 0.08 ) & & \\
			&&& && \\
			
			\multicolumn{6}{c}{Comparison of shared-path regimes}  \\
			&&& && \\
			$d(App, Relaxed)$ & ( 0.12, 0.28, 0.4, 0.2 ) & $d(App,  NC, Relaxed)$ & ( 0.12, 0.28, 0.4, 0.2 ) & 0.62 & ( 0.18, 1.06 ) \\
			& ( 0, 0.17, 0.17, 0.67 ) & & ( 0.25, 0.13, 0.38, 0.25 ) & & \\
			$d(App, Stringent)$ & ( 0.08, 0.15, 0.39, 0.39 ) & $d(App,  NC, Stringent)$ & ( 0.08, 0.15, 0.39, 0.39 ) & 1.08 & ( 0.32, 1.84 ) \\
			& ( 0.11, 0.56, 0.22, 0.11 ) & & ( 0.23, 0.23, 0.46, 0.08 ) & & \\
			\hline	
	\end{tabular}}
	\label{tab: belief}
\end{table}

\end{document}